\documentclass[reprint, showpacs, preprintnumbers, amsmath, amssymb, aps, nofootinbib]{revtex4}

\usepackage{color}
\usepackage{graphicx}
\usepackage{dcolumn}
\usepackage{overpic}
\usepackage{psfrag}
\usepackage{bm}

\usepackage{subfigure}

\def\OBJ{{\mathcal{J}}}
\def\LAG{{\mathcal{L}}}
\def\cc{\mathbf{c}}

\def\ha{{h^\dag}}

\def\qq{\mathbf{q}}
\def\nn{\mathbf{n}}
\def\xx{\mathbf{x}}

\def\cost{{\cos \theta}}
\def\cosb{{\cos \beta}}
\def\sint{{\sin \theta}}
\def\sinb{{\sin \beta}}
\def\SSS{{\left(\begin{array}{c}\sint \\ \sinb \end{array}\right)}}

\def\intt{{\int_{I}}}

\newcommand\be{\begin{equation}}
\newcommand\ee{\end{equation}}

\newcommand\Rey{\mbox{\textit{Re}}} 

\newcommand\Pe{\mbox{\textit{Pe}}}

\providecommand\bnabla{\boldsymbol{\nabla}}
\providecommand\bcdot{\boldsymbol{\cdot}}

\begin{document}


\title{Pancake making and surface coating: \\ optimal control of a gravity-driven liquid film}

\author{E. Boujo}
\affiliation{LadHyX, UMR CNRS 7646, Ecole Polytechnique, 91128 Palaiseau, France}

\author{M. Sellier}
\affiliation{Department of Mechanical Engineering, University of Canterbury. Private Bag 4800, Christchurch 8140, New Zealand
}

\date{June 2019}

\begin{abstract}
This paper investigates the flow of a solidifying liquid film on a solid surface subject to a complex kinematics, a process relevant to pancake making and surface coating. 
The flow is modeled using the lubrication approximation, with a gravity force whose magnitude and direction depend on the time-dependent orientation of the surface.
Solidification is modeled with a temperature-dependent viscosity.
Because the flow eventually ceases as the liquid film becomes very viscous, the key question this study aims to address is: what is the optimal surface kinematics for spreading the liquid layer uniformly? Two methods are proposed to tackle this problem. In the first one, the surface kinematics is assumed \textit{a priori} to be harmonic and parameterized. The optimal parameters are inferred using the Monte-Carlo method. This ``brute-force'' approach leads to a moderate improvement of the film uniformity compared to the reference case when no motion is imposed to the surface. The second method is formulated as an optimal control problem, constrained by the governing partial differential equation, and solved with an adjoint equation. Key benefits of this method are that no assumption is made on the form of the control, and that significant improvement in thickness uniformity are achieved with a comparatively smaller number of evaluations of the objective function.
\end{abstract}

\pacs{Valid PACS appear here}

\maketitle

\section{Introduction}
One of the  motivations for the work presented here is the process of cr\^epe making. 
Cr\^epe making involves pouring a fixed amount of batter on a hot pan, letting or forcing the batter to spread on the hot pan to obtain {\it optimal coverage}, and letting the batter cook. For cr\^epes, {\it optimal coverage} means uniformly thin, hole-free, and perfectly circular. Achieving this goal can however be quite challenging since as the batter spreads, it cooks at the same time and if the pan is left horizontal, the batter tends to solidify before reaching uniformly the rim of the pan. There are two main strategies to circumvent this issue. 
The first strategy involves using a blade to force spread a uniform layer of batter on the pan in a process reminiscent of blade coating. 
The other strategy involves tilting the pan in a swirling motion and forcing the batter to spread preferentially in the downslope direction of the span. As soon as this tilting is initiated, the axial-symmetry of the problem is broken and one cannot help but wonder what the optimal swirling motion is to achieve an optimally thin and round cr\^epe. This process has other important implications in chocolate manufacturing technologies, the coating of surfaces, or the production of thin elastic shells \cite{Lee16}. 

The key physical phenomena underpinning cr\^epe making involves the interaction of the liquid layer with the substrate kinematics and the solidification of the liquid layer. These are relevant in other related contexts of science and engineering. 
An archetypical illustration of the interaction between substrate kinematics and a liquid layer is spin coating for which a liquid is first deposited on a substrate which is then spun at high angular velocity to produce a thin liquid film which is subsequently cured, see for example \cite{Emslie58,Meyerhofer78,Sahu10}. In that context, the fluid is driven by the centrifugal forces and the substrate kinematics is a simple rotation about the vertical axis, possibly with variable angular velocity. In contrast, in the context of cr\^epe making, the batter is driven by gravity and the pan kinematics is much more complicated since it involves transient rotation around multiple axes. 
To the best of the authors' knowledge, modeling such flows has not been attempted before. Liquid layer solidification is relevant in a number of processes including coating \cite{Weinstein04}, ice accretion on surfaces \cite{Myers04,Moore17}, paint drying \cite{Schwartz10,Gaskell06}. 
The present work builds mostly on the literature related to geophysical flow solidification such as lava flows for which the viscosity is highly temperature dependent \cite{Bercovici94,Sansom04} and cooling is associated with eventual flow arrest. 

A key question the current contribution is aiming to address is what is the optimal pan kinematics to achieve a desired film profile? 
Such problems fall in the realm of optimal control in fluid mechanics for which substantial literature already exist, see \cite{Gunzburger02}, for example. Applications of optimal control in thin layer flows are reviewed in \cite{Sellier16}. 
Relevant contributions include that of Grigoriev where the author finds optimal heating strategies to actively suppress evaporative instabilities in thin liquid films \cite{Grigoriev02}, that of Sellier and Panda where the optimal substrate shape to control the free surface of a liquid film is inferred \cite{Sellier10}, or that of Papageorgiou et al. where the authors find the optimal source/sink distribution to control and stabilize falling liquid films \cite{Thompson16,Tomlin18}. 
Because thin liquid films are commonly described by the long-wave approximation which leads to  second-order or fourth-order parabolic partial differential equations (PDEs), depending on whether the effect of surface tension are prevalent or not, it is natural to apply the standard framework of optimal control of PDEs as described in \cite{Troltzsch10,Neitzel09}. 
This is what the present contribution proposes to do.

The paper is organized as follows.
Section~\ref{sec:model} describes the mathematical model for the spreading of a non-isothermal gravity current and the numerical solution strategy.
Results illustrating the gravity current dynamics (solution of the ``direct'' problem) are provided in Section~\ref{sec:results1}. 
The formulation of the optimal control problem (``adjoint'' problem and optimization) is given in Section~\ref{sec:opt_control_form} with the corresponding results in Section~\ref{sec:opt_control_res}.
The paper concludes with discussion and conclusions in Section~\ref{sec:disc_conc}.

\section{Description of the mathematical model}
\label{sec:model}

\subsection{Mathematical model}
\begin{figure}[htp]
\centering
\includegraphics[scale=0.78]{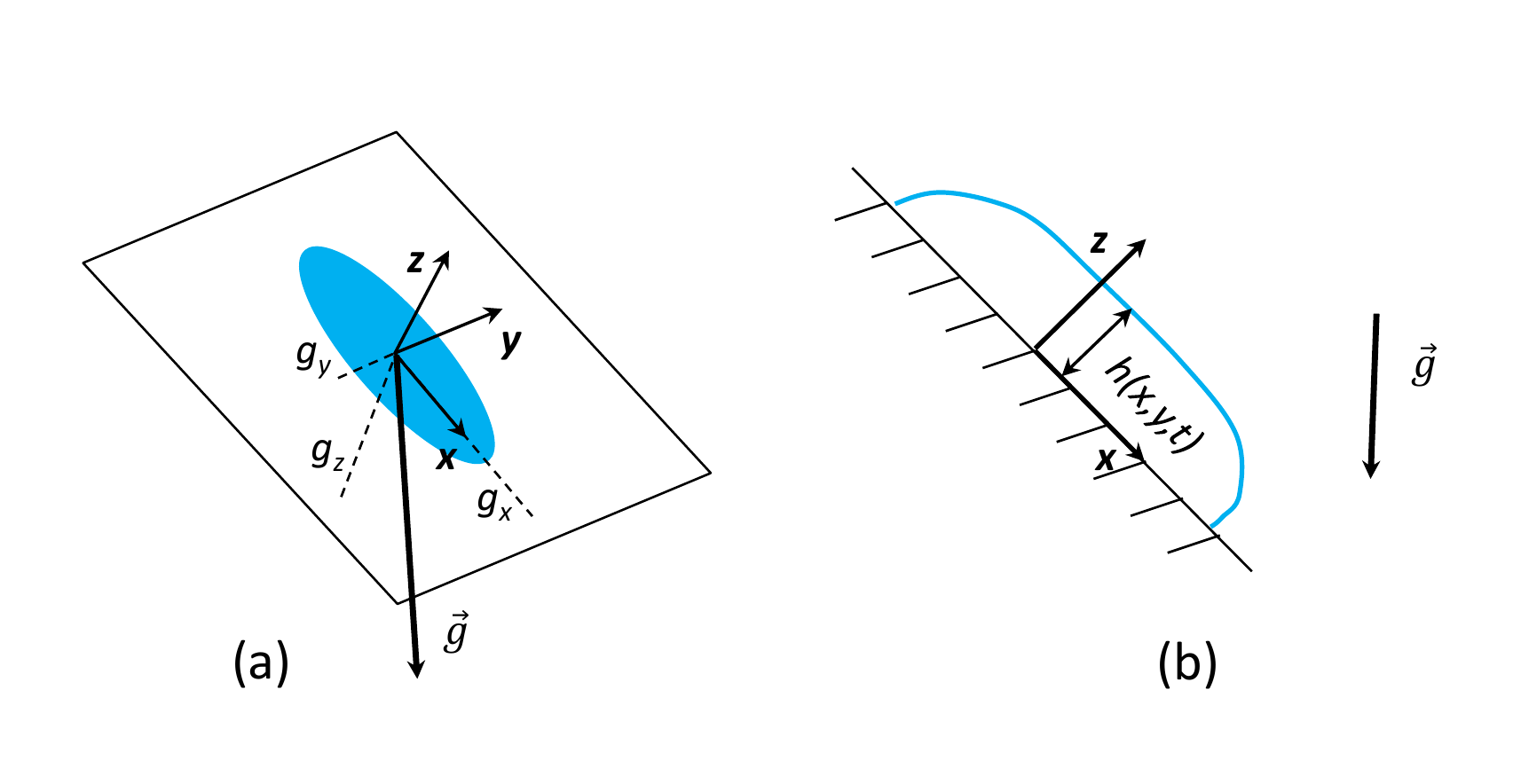}
\vspace{-1cm}
\caption{Schematic diagram of the viscous gravity current: (a) top view, (b) side view.}
\label{fig:sketch}
\end{figure}

The liquid layer is treated as a non-isothermal gravity current spreading over a surface whose inclination varies over time, see figure~\ref{fig:sketch}. 
In order to model a solidification process, the effect of the  phase transition on the dynamics
 is replaced by a temperature-dependent viscosity, and the layer is assumed to remain fluid at all times.  
Gravity currents have received considerable attention in the past and an excellent account of the current knowledge can be found in \cite{Leal10}. 
For an incompressible liquid film whose characteristic thickness is much smaller than its characteristic lateral dimensions $H/L=\epsilon\ll 1$ (lubrication theory) and whose motion at characteristic  velocity $U$ is of small Reynolds number $\Rey=\rho UL/\mu \ll 1$ (inertial effects much smaller than viscous effects),
the conservation equations for momentum, mass and energy reduce to:
\begin{eqnarray}
\frac{\partial }{\partial z} \left(\mu \frac{\partial u}{\partial z} \right) & = & \frac{\partial p}{\partial x}-\rho g_x \; , \label{eq:mom_x} \\
\frac{\partial }{\partial z} \left(\mu \frac{\partial v}{\partial z} \right) & = & \frac{\partial p}{\partial y}-\rho g_y \label{eq:mom_y} \; , \\
0 &=& \frac{\partial p}{\partial z}-\rho g_z  \label{eq:mom_z} \; ,\\
\frac{\partial u}{\partial x}+\frac{\partial v}{\partial y}+\frac{\partial w}{\partial z} & = & 0 \label{eq:continuity} \; , \\
\frac{d^2T}{dz^2} & = & 0 \label{eq:heat} \; ,
\end{eqnarray}
where $(u,v,w)$ are the components of the velocity vector, 
$p$ the pressure, 
$T$ the temperature, 
$\mu (T)$ the temperature-dependent dynamic viscosity,
$\rho$ the fluid density, and 
$(g_x,g_y,g_z)$ the components of the gravity acceleration vector. 
The coordinate system $(x,y,z)$ is attached to the substrate.
The liquid layer is bounded by the $xy$ plane from below and the free surface $h(x,y,t)$ from above. 

The substrate rotation is assumed to be sufficiently slow that the centrifugal force and Coriolis force may be neglected (this requires the characteristic rotation rate to be small compared to 
$\sqrt{\mu U/(\rho L H^2)}$
and to $\mu/(\rho H^2)$, respectively; we have verified that this is satisfied at almost all times for the substrate motions considered in this study; see Sec.~\ref{sec:results1}).

Eq.~(\ref{eq:heat}) is valid for thin films when unsteady thermal effects are negligible (such that the film is at thermal equilibrium in the vertical direction at all times) if the product of 
the square of the dimensionless film thickness $\epsilon^2$ with the P\'eclet number $\Pe=UL/\kappa$ (with $\kappa$ the thermal diffusivity) is small \cite{Sansom04}; we assume that this condition is satisfied.

The system depends on time via two effects. 
First,
as the substrate rotates, the orientation of the gravity vector changes with time. 
We represent the substrate kinematics by the two angles $\theta (t)$ and $\beta (t)$ such that $g_x = g \sin \theta$, $g_y = g \sin \beta$, and $g_z = g \cos \theta \cos \beta$. 
Second, (\ref{eq:heat}) is complemented with  time-dependent boundary conditions:
while the substrate is assumed to be at a fixed temperature $T_s$, the upper surface of the liquid layer is subject to convective heat transfer with the surrounding atmosphere at temperature $T_\infty$ 
(valid for small Biot numbers \cite{Oron1997}, which is satisfied for almost all the conditions considered in this study),
\be 
\left. 
 T(z=0)=T_s, 
\qquad
k \frac{\partial T}{\partial z} \right|_{z=h} = h_c\left( T_{\infty}-T(h)\right),
\label{heat_transfer}
\ee 
where $h_c$ is the convective heat transfer coefficient  and $k$ the liquid thermal conductivity.

As a result, the temperature-dependent viscosity of the liquid is also time-dependent.
We build on the analysis of \cite{Bercovici94,Sansom04} to account for the temperature-dependence of the viscosity in the lubrication approximation framework, and assume an exponential dependence of viscosity on temperature:
\be 
\mu(T) = \mu _0 e^{-\alpha T}.
\ee

The heat equation (\ref{eq:heat}) subject to the boundary conditions (\ref{heat_transfer}) can readily be integrated to yield an explicit relation between the temperature profile across the liquid layer $T(x,y,z,t)$ and its thickness $h(x,y,t)$. 
Accordingly
\begin{equation}
\label{eq:temp_dis}
T(x,y,z,t)=A'\frac{z}{h}+T_s \; ,
\quad 
\text{where } A'=\frac{h_c\left( T_{\infty}-T_s \right)}{\frac{k}{h}+h_c}. 
\end{equation}
This equation shows that the temperature distribution is ``enslaved'' to the film thickness distribution and the thermo-physical properties of the liquid.
Integrating equations (\ref{eq:mom_x})-(\ref{eq:mom_y}) with respect to $z$, subject to the no-slip boundary condition ($\mathbf{v}=\mathbf{0}$)
at the substrate and to the stress-free boundary condition at the free surface 
($-p \nn_s + \mu \nabla\mathbf{v} \cdot \nn_s=\mathbf{0}$, with $\nn_s$ the  unit vector normal to the surface, and neglecting capillary effects)
yields the following expressions for the velocity: 
\begin{eqnarray}
\mathbf{v}(x,y,z,t) & = & \frac{\rho g e^{\alpha T_s}}{\mu _0}\frac{h^2}{\left( \alpha A' \right)^2} \left( e^{\alpha A'\frac{z}{h}} \left( \alpha A' \frac{z}{h}-1-\alpha A'\right)+\alpha A'+1 \right) \mathbf{K} \; , \\
\mathbf{K}(x,y,t) & = &  \cos \theta(t) \cos \beta(t) \mathbf{\nabla} h -\begin{bmatrix} 
\sin \theta (t) \\
\sin \beta (t)
\end{bmatrix} \; .
\end{eqnarray}
The conservation of mass requires that 
\begin{equation}
\frac{\partial h}{\partial t}+\mathbf{\nabla}\cdot\mathbf{q}=0  \label{eq:governing} \; , 
\end{equation}
where the discharge $\mathbf{q}=\int\limits_{0}^{h}\mathbf{v} \,\text{d}z$ is given by
\begin{equation}
\mathbf{q} (x,y,t) = -\frac{\rho g e^{\alpha T_s}}{\mu _0}\frac{h^3}{\left( \alpha A' \right)^3} \left( 2e^{\alpha A'} -\left(\alpha A'\right)^2-2\alpha A'-2 \right) \mathbf{K} \; .
\label{eq:discharge}
\end{equation}
This second-order, non-linear, parabolic partial differential equation forms the basis of the forthcoming analysis. 
The computational domain $D$ is a disk of radius $R$ and the corresponding boundary is denoted by $\partial D$ with unit outward-pointing normal vector $\mathbf{n}_D$. 
On the boundary, a no-flux condition is imposed such that $\mathbf{q}\cdot\mathbf{n}_D=0$ on $\partial D$.

\subsection{Solution procedure}

Computing the evolution of the film thickness $h(x,y,t)$ consists in solving Eq.~(\ref{eq:governing}) for a prescribed substrate kinematics (prescribed $\beta (t)$ and $\theta (t)$) and given initial/boundary conditions.)
We call this problem the ``direct'' problem, as opposed to the ``adjoint'' problem involved in the optimal control (section~\ref{sec:opt_control_form}). 
In order to solve this non-linear partial differential equations, 
we use the COMSOL Multiphysics software which offers an integrated Finite Element environment to solve a wide range of PDEs. 
We use the \emph{Coefficient Form PDE} module which first requires casting the PDE in the following standard form
\begin{equation}
e_a \frac{\partial ^2 h}{\partial t^2}
+d_a \frac{\partial h}{\partial t}
+\nabla \cdot \left( -c \nabla h -\mathbf{\alpha} h + \mathbf{\gamma} \right) + \mathbf{\beta}\cdot\nabla h + a h = f 
\; ,
\label{eq:comsolPDE}
\end{equation}
where the coefficients $e_a$, $d_a$, $c$, $\mathbf{\alpha}$, $\mathbf{\beta}$, $a$, $\mathbf{\gamma}$, $f$ 
to be specified are identified from (\ref{eq:governing}).
The computational domain $D$ is meshed with approximately 9,200 Lagrange quadratic elements. 
The variable-order, variable-step-size backward differentiation formula (BDF) is used for time integration and the time step is limited to $10^{-2}$~s. 
These settings ensure a robust, mesh-independent solution.

\section{Direct problem results}
\label{sec:results1}

The goal of the present work is to identify the optimal substrate kinematics to achieve a uniform coverage by the liquid layer of the substrate. 
Because the substrate is a disk of radius $R$, the optimal layer thickness is 
$h_\text{opt} = V_D/(\pi R^2) $
for a given volume of liquid $V_D$ spread uniformly over the whole disk.
Therefore, a good measure of thickness uniformity is
\begin{equation}
\label{eq:OF}
\mathcal{U}(t) = \iint_{D}\left( h(x,y,t)-h_\text{opt} \right)^2 \text{d}x \text{d}y.
\end{equation}
A uniform film yields $\mathcal{U}=0$, while large deviations from $h_\text{opt}$ yield large values of $\mathcal{U}$.
The question therefore arises as to how the various physical parameters involved in the problem affect the film thickness uniformity. The next two subsections illustrate:
\begin{itemize}
\item 
first, the effect of the heat transfer and the resulting viscosity variation on the levelling dynamics when the substrate is held fixed and horizontal;
\item 
second, the effect of the substrate kinematics on the film thickness uniformity for prescribed heat transfer conditions. 
\end{itemize}
The parameter space is vast, but to fix ideas and for illustration purposes, we set the following parameters inspired by the process of cr\^epe making (see also Table~\ref{Tab:param}).

\begin{table}
\begin{center}
  \begin{tabular}{ | c | c | c | c | c | c |}
    \hline
    Parameter & $\rho$ & $k$ & $\mu_0$ & $\alpha$ & $R$\\
    &  (kg m$^{-3}$)  &  (W m$^{-1}$ K$^{-1}$)  &  (Pa~s)  & ($^o$C$^{-1}$) & (m)  \\
    Value & 1000 & 1 & $1.668$ & $2.56\times 10^{-2}$ & $1.5\times 10^{-1}$\\
    \hline
    Parameter & $V_D$ & $h_c$ & $T_\infty$ & $T_s$ & $h_\text{opt}=V_D/(\pi R^2)$ \\
              & (m$^3$) &  (W m$^{-2}$ K$^{-1}$)  & ($^o$C) & ($^o$C) & (m) \\
    Value & $1.414\times 10^{-4}$ & variable & 200 & 20 & $2\times10^{-3}$ \\
    \hline
  \end{tabular}
\caption{Parameter table.}
\label{Tab:param}
\end{center}
\end{table}

The viscosity parameters $\mu _0$ and $\alpha$ are chosen such that $\mu (T=200^o$C$)=5\times10^{-3}$~Pa~s and $\mu (T=20^o$C$)=10$~Pa~s. 
The thermo-physical properties are such that the liquid layer cools down from a temperature close to the surrounding temperature $T_\infty$ at $t=0$, down to the substrate temperature $T_s$ as $t\rightarrow t_f$. 
The volume is chosen such that $h_\text{opt}=2\times 10^{-3}$~m. 
The initial condition is a liquid column of radius $R_i = R/3$ and thickness $h_i$ required to match the volume constraint. 
This liquid column rests on a precursor film of thickness $h_i/100$ to regularize the degenerate PDE, and the initial thickness $h(x,y,t=0)$ is smoothed over a narrow band such that the profile is twice differentiable to ease convergence without the need for an exceedingly fine mesh.

As mentioned in Sec.~\ref{sec:model}, the centrifugal and Coriolis forces can be neglected if the  characteristic rotation rate $\Omega$ is small compared to  $\Omega_{1} = \sqrt{\mu U/(\rho L H^2)}$
and $\Omega_{2} = \mu/(\rho H^2)$, respectively. 
We note that, because the film thickness, fluid velocity and viscosity vary in space and time, so do $\Omega_{1}$ and $\Omega_{2}$.
Given the above choice of parameters, typical values may be taken in the range 
$H \sim [2, 5]\times 10^{-3}$~m,
$U \sim [2, 5]\times 10^{-2}$~m~s$^{-1}$, and
$\mu \sim [0.1, 1]$~Pa~s (corresponding to $T \sim [20, 100]^o$C);
the typical lateral extension is taken as $L \sim 0.15$~m.
This yields 
$\Omega_{1} \sim [0.8, 9]$~rad~s$^{-1}$ and
$\Omega_{2} \sim [5, 250]$~rad~s$^{-1}$.
In the following, the control will impose rotation rates of the order of $\Omega \sim 0.5$~rad~s$^{-1}$ (see e.g. Figs.~\ref{fig:MC_controls} and \ref{fig:curves_opt_ctrl_adjoint}), generally smaller than both $\Omega_{1}$ and $\Omega_{2}$. 
Therefore, at the exception of short times when $\mu$ and $U$ are small and $H$ is large, 
the centrifugal and Coriolis forces can be reasonably neglected.

\subsection{Effect of heat transfer}
\label{sec:fixed_substrate}

First, the substrate is held fixed and horizontal. 
The fluid layer, initially at high-temperature, has a low viscosity and therefore spreads well on the substrate. 
As the fluid spreads and cools down, its viscosity increases, resulting in a slowdown of the spreading. 
From this reasoning, it is clear that the cooling rate plays a major role on the spreading dynamics. 
Two processes are at play: 
(i) heat convection with the surrounding atmosphere at high temperature $T_\infty$, and 
(ii) cooling from the substrate at low temperature $T_s$. 
The weaker the heat convection, the faster the cooling. 
Therefore, as $h_c$ is reduced, cooling occurs faster, slowing the spreading and reducing the substrate coverage.

\begin{figure}[htp]
\begin{center}
\subfigure[]{
\label{fig:FS_prof_h3}
  \includegraphics[width=0.45\textwidth, trim={35mm 90mm 40mm 90mm}, clip]{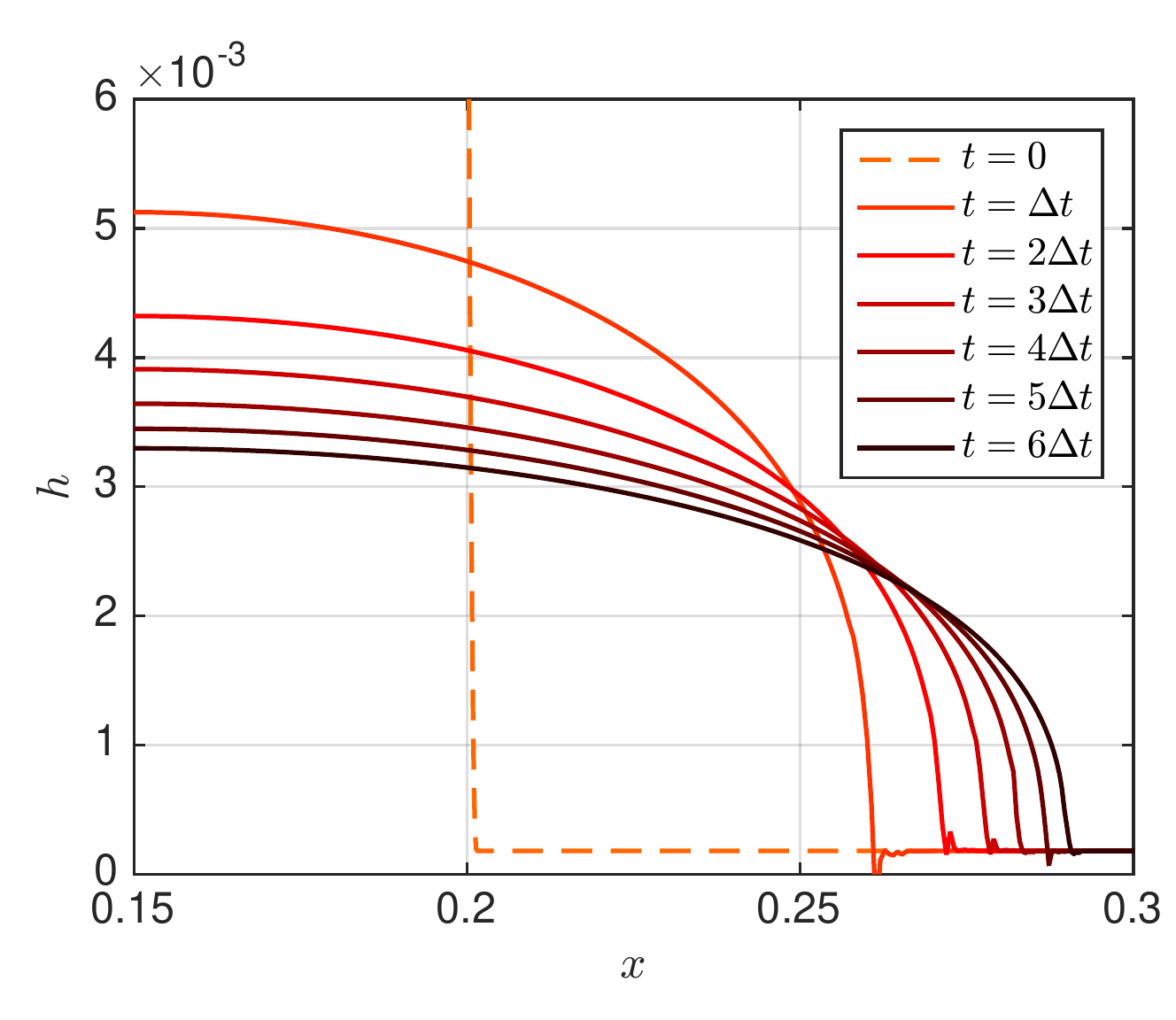}
  }
\subfigure[]{
\label{fig:FS_prof_h300}
  \includegraphics[width=0.45\textwidth, trim={35mm 90mm 40mm 90mm}, clip]{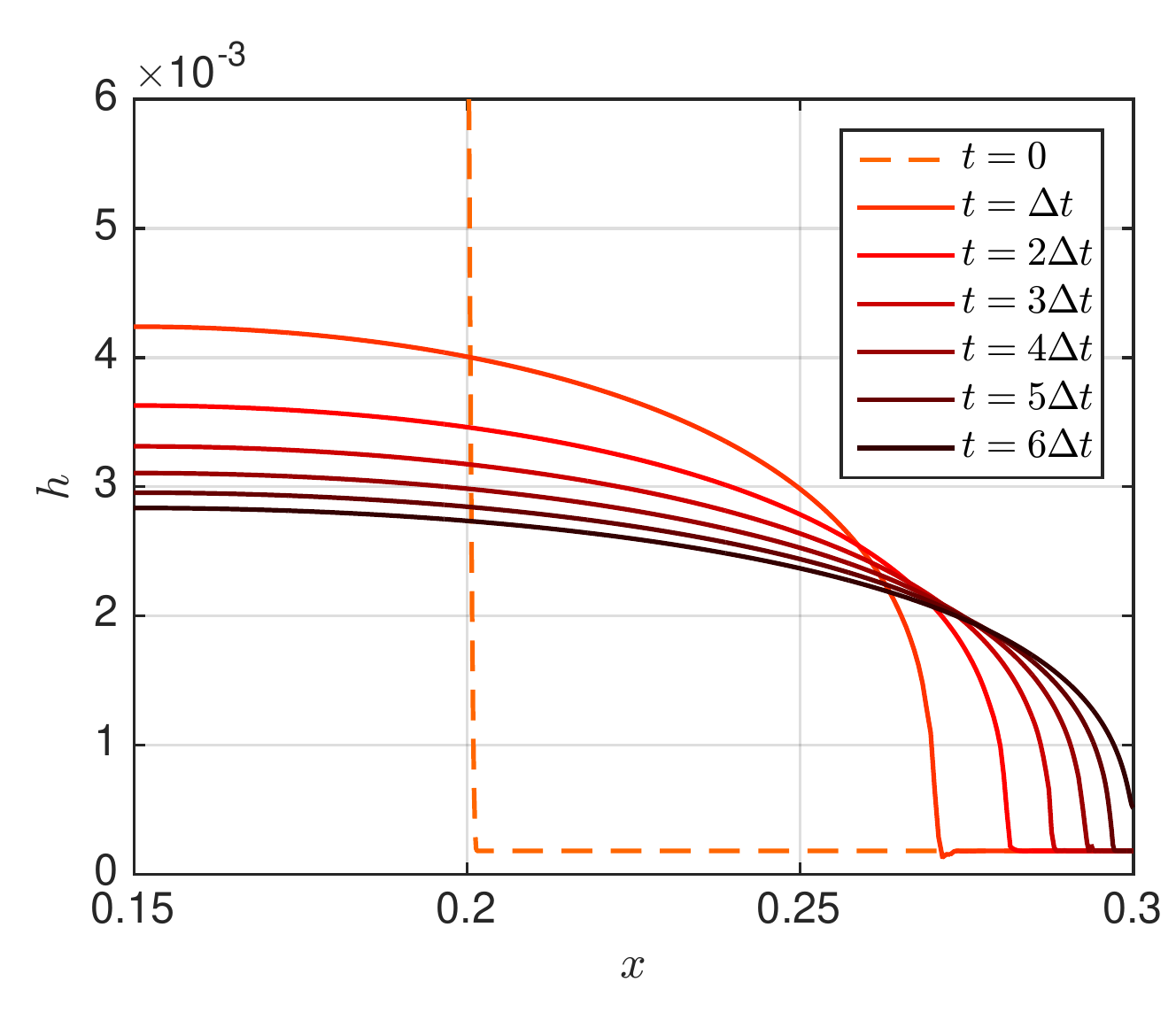}
  }
\caption{Free surface profile $h(x)$ along $y=0$ as a function of time:
(a)~$h_c=3$~W~m$^{-2}$~K$^{-1}$; 
(b)~$h_c=300$~W~m$^{-2}$~K$^{-1}$. 
The profile is shown for $x\in [R,2R]$ because the flow is axisymmetric. 
The gravity current spreads outwards and profiles are $\Delta t=4.29$~s apart until $t_f$ = 30~s.
}
\label{fig:FS_evol}
\end{center}
\end{figure}

Fig.~\ref{fig:FS_evol} shows the evolution of the free surface  for a high value of $h_c$ for which cooling is slow (Fig. \ref{fig:FS_prof_h300}) and a low value of $h_c$ for which cooling is fast (Fig.~\ref{fig:FS_prof_h3}). 
The advancing front is clearly shown to progress faster for the higher value of $h_c$ reaching the domain boundary at the latest times. 
Consequently, the leveling is better for that case, as one would expect.

\begin{figure}[htp]
\begin{center}
   \subfigure[]{
   \label{fig:FS_against_h}
   \includegraphics[width=0.45\textwidth, trim={33mm 90mm 40mm 92mm}, clip]{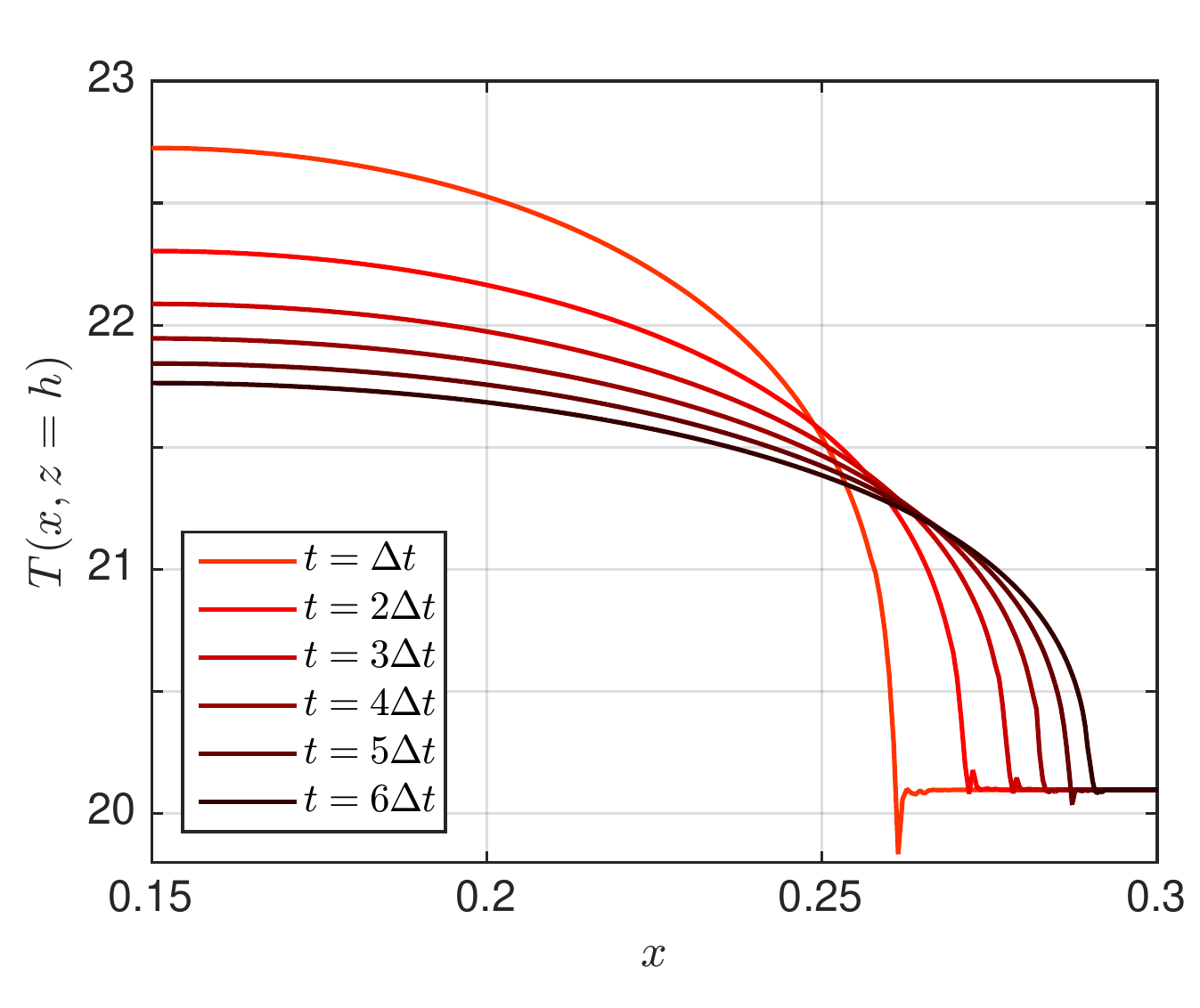}
   }
   \subfigure[]{
   \label{fig:temp_against_h}
   \includegraphics[width=0.45\textwidth, trim={33mm 90mm 40mm 92mm}, clip]{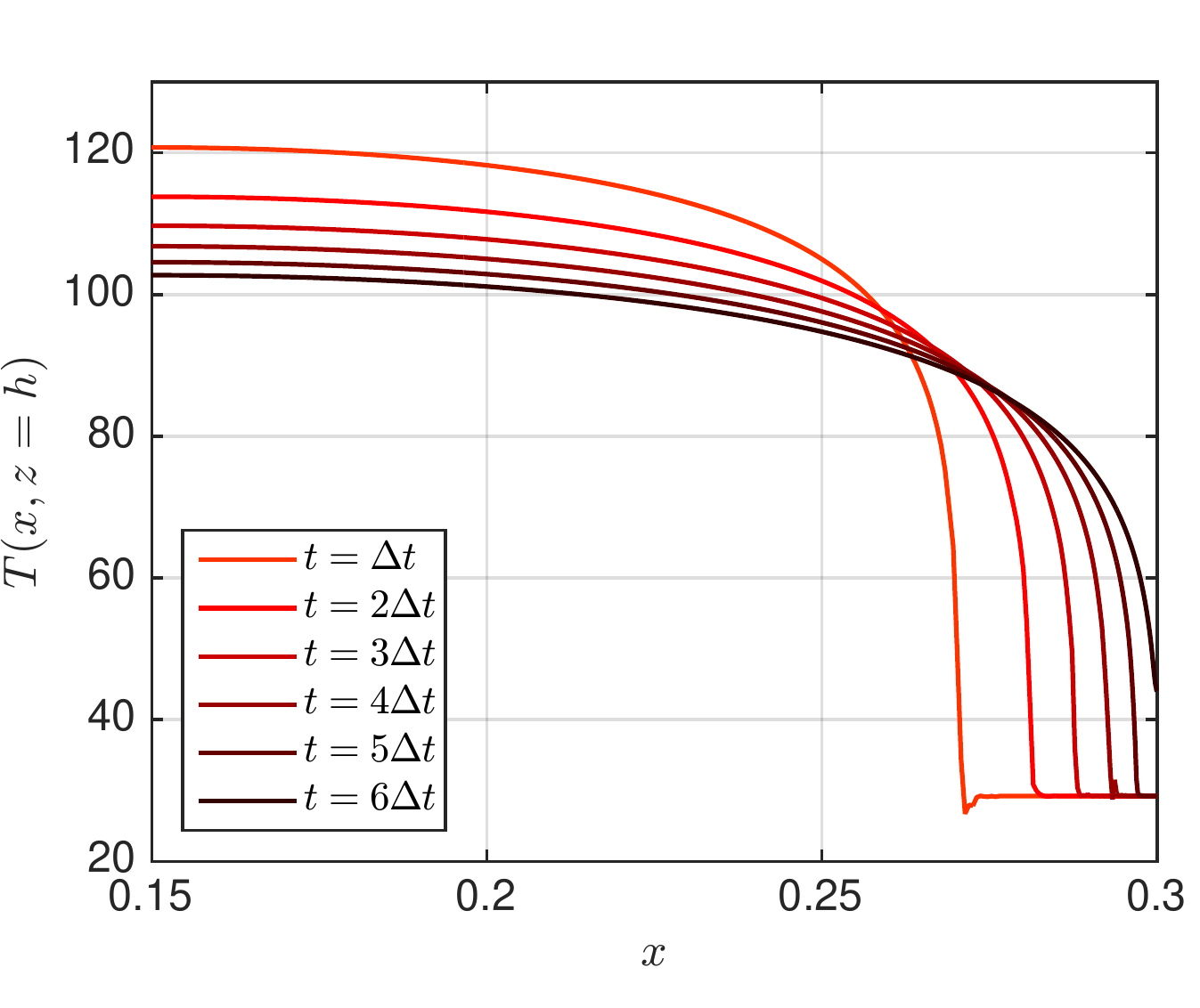}
   }
\caption{
Effect of heat transfer coefficient on the free surface temperature distribution $T(x,z=h)$ along $y=0$: 
(a)~$h_c=3$~W~m$^{-2}$~K$^{-1}$; 
(b)~$h_c=300$~W~m$^{-2}$~K$^{-1}$. 
Profiles are $\Delta t=4.29$~s apart until $t_f$ = 30~s.
}
\end{center}
\label{fig:against_h}
\end{figure}

Fig.~\ref{fig:against_h} confirms that the free surface temperature calculated according to Eq.~(\ref{eq:temp_dis}) decreases faster for 
$h_c=3$~W~m$^{-2}$~K$^{-1}$ than for 
$h_c=300$~W~m$^{-2}$~K$^{-1}$, and therefore the film temperature gets closer to the surface temperature $T_s$. This faster cooling corresponds to a faster increase of the viscosity.

\begin{figure}[htp]
\centering
\subfigure[]{
\includegraphics[trim={8mm 65mm 22mm 70mm}, clip, height=5.8cm]{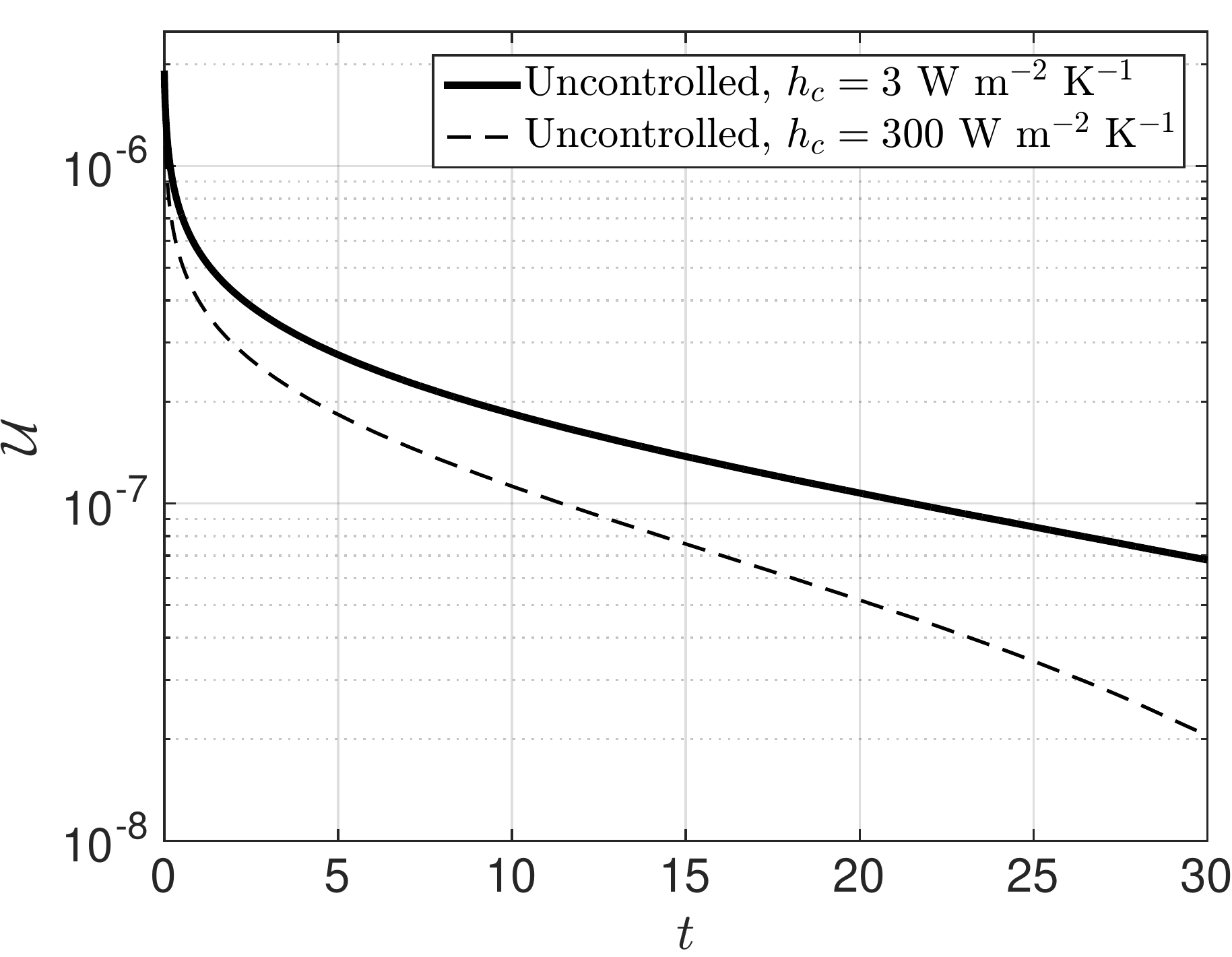}
}
\subfigure[]{
\includegraphics[trim={8mm 65mm 22mm 70mm}, clip, height=5.8cm]{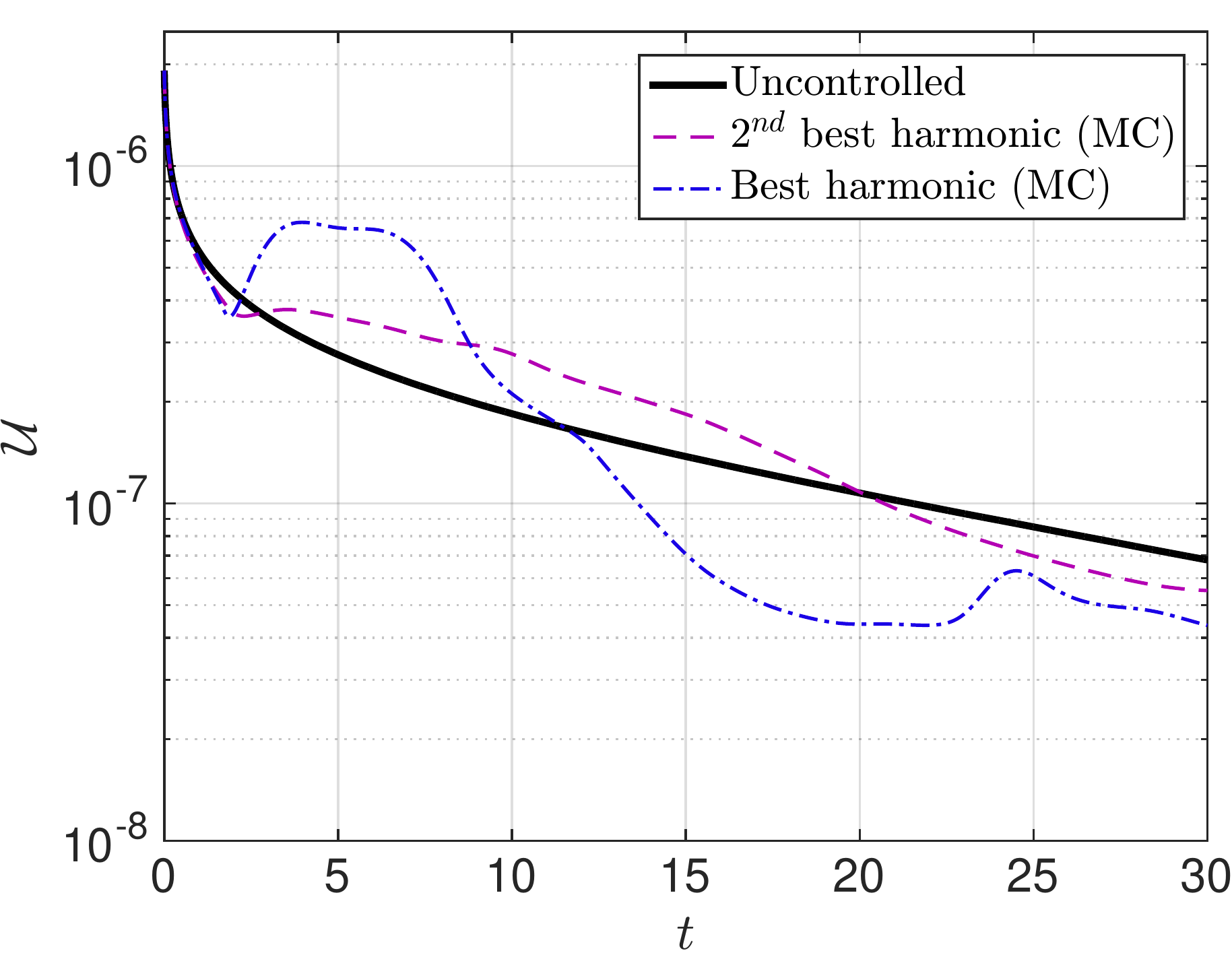}
}
\caption{
Time evolution of the measure $\mathcal{U}(t)$ of film thickness uniformity (see Eq.~(\ref{eq:OF})).
(a)~No control, $h_c=3$ and  $300$~W~m$^{-2}$~K$^{-1}$;
(b)~No control vs. best harmonic controls of the form~(\ref{eq:time_harm_1})-(\ref{eq:time_harm_2}) obtained with a Monte-Carlo algorithm ($h_c=3$~W~m$^{-2}$~K$^{-1}$).
}
\label{fig:heffectonOF}
\end{figure}

The effect of $h_c$ on the layer uniformity (\ref{eq:OF}) is shown in Fig.~\ref{fig:heffectonOF}. 
At $t=30$~s, $\mathcal{U}(t)$ has decreased down 
to $7.1\times 10^{-8}$~m$^4$ for
$h_c=3$~W~m$^{-2}$~K$^{-1}$, and 
to $2.3\times 10^{-8}$~m$^4$ for
$h_c=300$~W~m$^{-2}$~K$^{-1}$. 
This confirms the slower leveling for smaller values of the convective heat transfer coefficient.

In the remainder of this study, the heat transfer coefficient is set to $h_c=3$~W~m$^{-2}$~K$^{-1}$. 
This comparatively low value, associated with a fast cooling and poor leveling by gravity  when the substrate is fixed, calls for a suitable substrate motion in order to spread the film more uniformly.

\subsection{Effect of substrate kinematics}
\label{sec:Monte_Carlo}

\begin{figure}[htp]
\begin{center}
\subfigure[]{
\label{fig:MC1_controls}
  \includegraphics[trim={10mm 65mm 22mm 70mm}, clip, height=5.8cm]{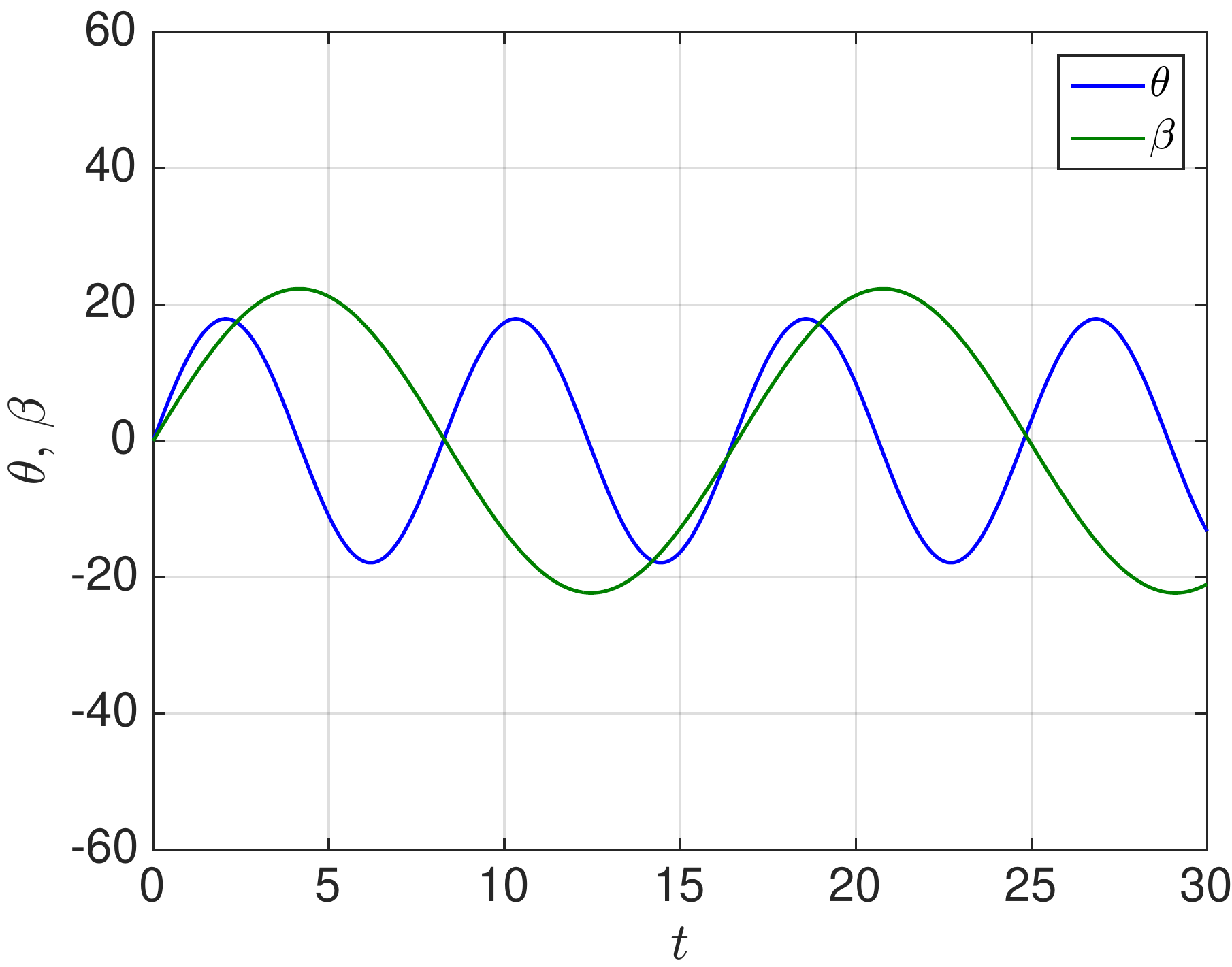} 
}
\subfigure[]{
\label{fig:MC2_controls}
  \includegraphics[trim={10mm 65mm 22mm 70mm}, clip, height=5.8cm]{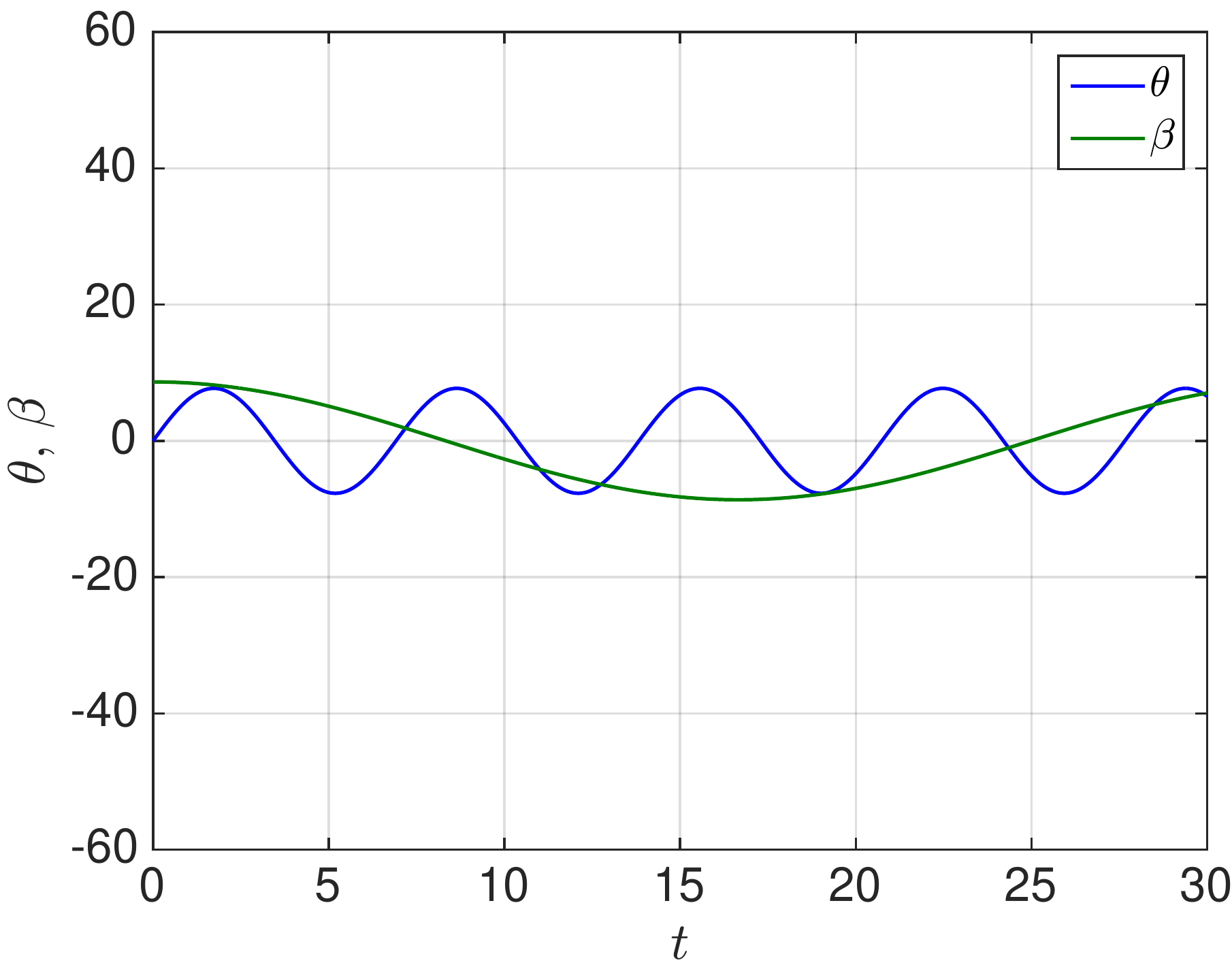} 
}
\caption{
Optimal harmonic controls minimizing $\mathcal{U}(t_f)$  with the Monte-Carlo method: 
(a)~best control,
(b)~second best control 
(see definitions and numerical values in the text).
}
\label{fig:MC_controls}
\end{center}
\end{figure}

We now investigate the effect of the substrate motion. 
Without any \textit{a priori} knowledge of what kind of kinematics can efficiently spread the film, we choose two families of time-harmonic angle laws,
\be 
\theta(t) = A_1 \sin\left( \frac{2\pi}{T_1} t \right), \quad
\beta(t) = A_2 \sin\left( \frac{2\pi}{T_2} t \right),
\label{eq:time_harm_1}
\ee
and
\be 
\theta(t) = A_1 \sin\left( \frac{2\pi}{T_1} t \right), \quad
\beta(t) =
 A_2 \cos\left( \frac{2\pi}{T_2} t \right),
\label{eq:time_harm_2}
\ee
and perform a broad exploration of the space parameter $(A_1, A_2, T_1, T_2)$.
A Monte-Carlo algorithm picks random combinations of amplitudes $A_1$, $A_2$  defined 
in $[5, 45]^o$  and periods $T_1$, $T_2$ defined 
in $[5, 50]$~s; for each combination, it then solves the governing equations over $t \in [0,30]$~s and evaluates the final uniformity $\mathcal{U}(t_f)$.
The best results obtained after 1000 evaluations with (\ref{eq:time_harm_1}) and 1000 evaluations with (\ref{eq:time_harm_2}) are shown in Fig.~\ref{fig:MC_controls}.

From the set of Monte-Carlo realizations, the best harmonic kinematics minimizing $\mathcal{U}(t_f)$ corresponds to Eq.~(\ref{eq:time_harm_1}) with $A_1$=17.9$^o$, 
$A_2$=22.3$^o$, 
$T_1$=8.26~s, 
$T_2$=16.6~s, see Fig. \ref{fig:MC1_controls}. 
For comparison, the second best kinematics obtained with 
$A_1$=7.70$^o$, 
$A_2$=8.65$^o$, 
$T_1$=6.91~s, 
$T_2$=33.3~s in Eq.~(\ref{eq:time_harm_2}) is shown in Fig.~\ref{fig:MC2_controls}. 
The value of $\mathcal{U}(t_f)$ is $4.19\times 10^{-8}$ m$^4$ and $5.69\times 10^{-8}$ m$^4$ for the best and second best control, respectively. 
This moderate improvement in film uniformity is confirmed on Fig.~\ref{fig:heffectonOF} showing the time evolution of $\mathcal{U}(t)$. 
For both cases, we note that the amplitude of the variations in $\theta$ and $\beta$ are of the same order. Moreover, both signals are out of phase and the longest period is an approximate multiple of the shortest one (the period of $\beta$ is approximately twice that of $\theta$ for the best harmonic control, and four times for the second best). 
In the context of cr\^epe making, this kinematic corresponds to a slow rocking motion in one direction composed with a transverse rocking motion at a higher frequency.

\begin{figure}
\centering
\includegraphics[scale=1]{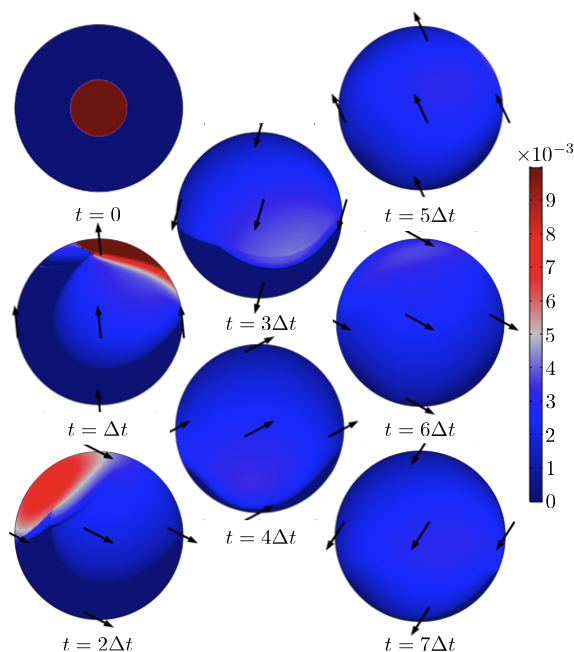}
\caption{
Contours of film thickness $h(\xx,t)$ for the optimal harmonic kinematics minimizing 
$\mathcal{U}(t_f)$, 
obtained with the Monte-Carlo method.  
The optimal parameters in Eq.~(\ref{eq:time_harm_1}) are 
$A_1$=17.9$^o$, 
$A_2$=22.3$^o$, 
$T_1$=8.26~s, 
$T_2$=16.6~s.
The time interval between snapshots is $\Delta t=4.29$~s. 
Arrows represent the direction of the projection of the gravity vector on the surface plane. 
See video 1 in the supplementary material \cite{SM}.}
\label{fig:MC1}
\end{figure}

\begin{figure}
\centering
\includegraphics[scale=1]{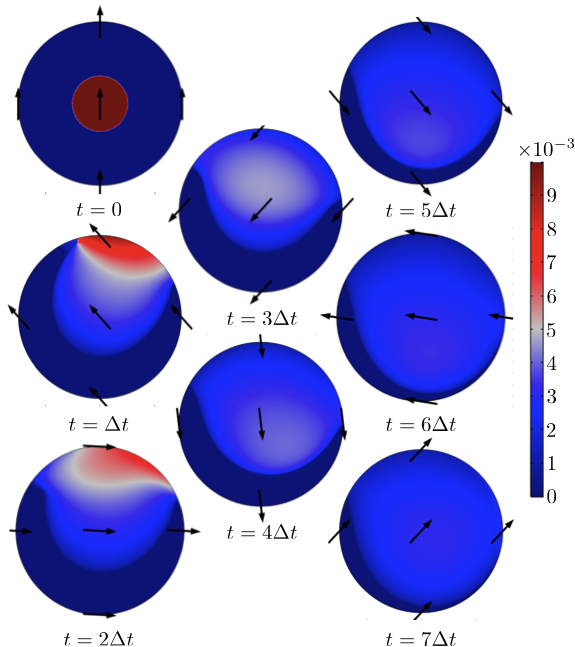}
\caption{Contours of film thickness $h(\xx,t)$ for the second best harmonic kinematics minimizing $\mathcal{U}(t_f)$, obtained with the Monte-Carlo method.  
The optimal parameters in Eq.~(\ref{eq:time_harm_2}) are 
$A_1$=7.70$^o$,
$A_2$=8.65$^o$, 
$T_1$=6.91~s, 
$T_2$=33.3~s.
Same representation as Fig.~\ref{fig:MC1}.
See video 2 in the supplementary material \cite{SM}.}
\label{fig:MC2}
\end{figure}

The corresponding evolution of the film thickness distribution for the best and second best harmonic controls is illustrated in Figs.~\ref{fig:MC1} and \ref{fig:MC2}, respectively. 
In order to better grasp the substrate kinematics on these figures, the orientation of the substrate is indicated by a vector in the direction of the  projection of the gravity vector in the plane of the substrate. 
Videos are also available as supplementary material \cite{SM}.
In both figures, a positive $\beta$ corresponds to an upwards pointing arrow and a positive $\theta$ to a rightwards pointing arrow. 
These figures and videos suggest that a good strategy involves draining the fluid to one end of the surface (low-frequency rocking around the first axis) and redistributing it by a sideways tilt of the surface (higher-frequency rocking around a second, perpendicular, axis).

\section{Optimal control formulation}
\label{sec:opt_control_form}

The previous section illustrated the spreading 
behavior of the fluid on a substrate at rest (sec. \ref{sec:fixed_substrate}), and on a substrate moving with time-harmonic kinematics (sec. \ref{sec:Monte_Carlo}).
We now turn our attention to the question of finding better kinematics in order to improve the uniformity of the film. 
In this section we present the adjoint framework used to modify the kinematics  so as to improve uniformity iteratively, in an efficient way.

\subsection{Optimization problem}

Let us define our objective:
we wish to find a control law $\cc(t)=(\theta(t), \beta(t))$ over the time interval $I=[0,t_f]$ that minimizes a measure of the difference between the final film thickness and the ideal uniform  thickness
$h_\text{opt} = V_D/(\pi R^2)$.
We may wish to achieve this at minimal control cost, i.e. by spending as little energy as possible to move the substrate. 
Therefore, we introduce a scalar objective function that includes both the final uniformity and the total cost of the control
\begin{align}
\OBJ_{t} (h,\cc) &= \iint_D (h(\xx,t_f)-h_\text{opt})^2 \, \mathrm{d}\xx
+ \dfrac{\gamma}{t_f} \intt \left( \theta(t)^2+\beta(t)^2 \right) \,\mathrm{d}t
\nonumber
\\
&= \mathcal{U} (t_f) + \dfrac{\gamma}{t_f} \mathcal{C},
\label{eq:J1_J2}
\end{align}
where the weight $\gamma \geq 0$ can be seen as the unit cost of the control and allows one to choose how much the control should be penalized.
The subscript \textit{t} in $\OBJ_{t}$ stands for ``terminal control'' because the uniformity is evaluated at the final time $t_f$ only (see Section~\ref{sec:regulation} for ``regulation control'', where the uniformity is evaluated at all times).
The optimization problem reads
\begin{align}
\min_{\cc(t)} \OBJ_{t}(h,\cc)
\quad 
\mbox{ subject to (\ref{eq:governing}),}
\label{eq:constrained_opt_pb}
\end{align}
together with the initial and boundary conditions specified in sections~\ref{sec:model}-\ref{sec:results1}.

This constrained optimization problem can be solved with gradient-based methods, which requires  computing  the gradient $\mathrm{d} \OBJ_{t}/\mathrm{d} \cc$ of the objective function with respect to the control. 
Given the infinite number of degrees of freedom in the control $\cc(t)$ (which becomes a finite but  large number after discretization), evaluating the gradient numerically with a finite-difference approach (where the value  of $\OBJ_{t}$ is evaluated repeatedly  for each degree of freedom perturbed once at a time) has a  prohibitive computational cost. 
Instead, as classically done in optimal control problems, one can use an adjoint-based approach and compute the gradient very efficiently.

\subsection{Adjoint equation}

Computing the gradient $\mathrm{d} \OBJ_{t}/\mathrm{d} \cc$ efficiently relies on solving an adjoint equation. Let us detail now the derivation of this equation.
We transform the constrained problem (\ref{eq:constrained_opt_pb}) into an unconstrained problem by introducing the  
Lagrangian
\begin{align}
\LAG_t(h,\cc,\ha)
=&
\OBJ_{t}(h,\cc) 
- 
\intt \iint_D \ha \left(\dfrac{\partial h}{\partial t} + \bnabla \bcdot \qq \right) \,\mathrm{d}\xx \,\mathrm{d}t
\\
=&
\iint_D (h(\xx,t_f)-h_\text{opt})^2 \, \mathrm{d}\xx
+ \dfrac{\gamma}{t_f} \intt \left( \theta(t)^2+\beta(t)^2 \right) \,\mathrm{d}t
\nonumber
\\
& 
-\intt \iint_D \ha \left(\dfrac{\partial h}{\partial t} + \bnabla \bcdot \qq \right) \,\mathrm{d}\xx \,\mathrm{d}t,
\end{align}
where the adjoint variable $\ha(\xx,t)$  is a Lagrange multiplier
that
enforces the governing equation (\ref{eq:governing}).
In the following, derivatives on functional spaces are understood as Fr\'echet derivatives, which we denote loosely for any scalar or vectorial quantity $s$:
\be 
\dfrac{\partial f(s)}{\partial s} \delta s
=
\lim_{\epsilon \to 0}
\frac{f(s + \epsilon \delta s) - f(s)}{\epsilon} \quad \forall \, \delta s.
\ee
The total derivative of the Lagrangian with respect to the control reads
\begin{align}
\dfrac{\mathrm{d}\LAG_t}{\mathrm{d}\cc}
=
\dfrac{\partial\LAG_t}{\partial\cc}
+
\dfrac{\partial\LAG_t}{\partial h}
\dfrac{\mathrm{d} h}{\mathrm{d}\cc}
+
\dfrac{\partial\LAG_t}{\partial \ha}
\dfrac{\mathrm{d} \ha}{\mathrm{d}\cc},
\label{eq:Lagr_tot_der}
\end{align}
and it reduces to
\begin{align}
\dfrac{\mathrm{d}\LAG_t}{\mathrm{d}\cc}
=
\dfrac{\partial\LAG_t}{\partial\cc}
\end{align}
when both 
$\partial\LAG_t/\partial h=0$ and 
$\partial\LAG_t/\partial \ha=0$.
If, in addition, the governing equation (\ref{eq:governing}) is satisfied, then
 $\LAG_t = \OBJ_{t}$ by construction, and thus the gradient of interest is obtained as
\begin{align}
\dfrac{\mathrm{d} \OBJ_{t}}{\mathrm{d}\cc}
=
\dfrac{\partial\LAG_t}{\partial\cc}.
\end{align}
Specifically, the gradient reads
\begin{align}
\dfrac{\mathrm{d} \OBJ_{t}}{\mathrm{d}\cc}
&= 
\dfrac{2\gamma}{t_f} 
\left(
\begin{array}{c} \theta(t) \\ \beta(t) 
\end{array} \right)
- \iint_D \ha 
\left(
\dfrac{\partial \left(\bnabla \bcdot \qq\right)}{\partial \theta},
\dfrac{\partial \left(\bnabla \bcdot \qq\right)}{\partial \beta} 
\right)^T
\,\mathrm{d}\xx.
\label{eq:dJdc}
\end{align}
It has two time-dependent components, that will be used to update the control $(\theta(t), \beta(t))$.

The yet unknown adjoint variable $\ha(\xx,t)$ remains to be specified via the condition
$\partial\LAG_t/\partial h=0$, 
 which is discussed below. 
Note that the other condition to be satisfied, 
$\partial\LAG_t/\partial \ha=0$,
is by construction the governing equation (\ref{eq:governing}).

At this point, we  note that one can  simplify the problem to a large extent when $\alpha A' \ll 1$, which is well verified in practice for our choice of parameters $\alpha$, $k$, $h_c$ and $T_\infty-T_s$.
In this case, the discharge (\ref{eq:discharge}) is well approximated by
\begin{align}
\qq \simeq -\dfrac{e^{\alpha T_s} \rho g}{3 \mu_0}  h ^3
\left( \cost \cosb \bnabla h - \SSS \right).
\end{align}
This expression will be used to derive an approximate adjoint equation, and therefore an approximate gradient. 
Note however that the direct equation for $h(\xx,t)$ is  solved in its full form  (\ref{eq:governing}).
For simplicity, in the following we denote  $W = e^{\alpha T_s} \rho g / (3 \mu_0 )$,
such that the vector in the second term of (\ref{eq:dJdc}) reads
\begin{align}
 \dfrac{\partial \left(\bnabla \bcdot \qq\right)}{\partial \theta}
=
W \bnabla \bcdot \left[  h ^3
\left( \sint \cosb \bnabla h 
+
\left(\begin{array}{c}\cost \\ 0 \end{array}\right)
 \right) \right],
\label{eq:J1}
\\
 \dfrac{\partial\left(\bnabla \bcdot \qq\right)}{\partial \beta}
=
W  \bnabla \bcdot  \left[ h ^3
\left( \cost \sinb \bnabla h 
+
\left(\begin{array}{c} 0 \\ \cosb \end{array}\right)
 \right) \right].
\label{eq:J2}
\end{align}

The condition 
 $\partial\LAG_t/\partial h=0$ yields, after integration by parts, the adjoint equation associated with this terminal control problem,
\begin{align}
& \dfrac{\partial \ha}{\partial t}
+ W  
\bnabla  \bcdot \left( h^3  \cost \cosb \bnabla \ha  \right) 
-3 W h^2 
 \left( \cost \cosb \bnabla h - \SSS \right) 
 \bcdot \bnabla\ha
= 0,
\label{eq:adj}
\end{align}
together with terminal and boundary conditions for $\ha(\xx,t)$:
\begin{align}
& \ha(\xx,t_f) = 2 (h(\xx,t_f)-h_\text{opt}),
\label{eq:adj_init}
\\
& \bnabla \ha \bcdot \nn_D = 0 
\quad \mbox{ on }\partial D.
\label{eq:adj_BC}
\end{align}
Given the terminal condition (\ref{eq:adj_init}), the adjoint equation (\ref{eq:adj})  must be solved backward in time from $t=t_f$ to $t=0$. Note  that the adjoint equation is linear in $\ha(\xx,t)$, and its coefficients depend on the direct solution $h(\xx,t)$.

\subsection{Solution procedure}
\label{sec:adjoint_sol_proc}

The iterative optimization procedure is the following:
\begin{enumerate}
\item
Given a tentative control $\cc_k(t)=(\theta_k(t),\beta_k(t))$, solve the governing equation (\ref{eq:governing})
$\rightarrow$ obtain the solution $h(\xx,t)$;
\item
Given the control $\cc_k(t)$, and the solution $h(\xx,t)$ from step (1),
solve the adjoint equation (\ref{eq:adj})
$\rightarrow$ obtain the adjoint solution $\ha(\xx,t)$;
\item
Given the  control  $\cc(t)$,  the solution $h(\xx,t)$  from step (1) and the adjoint solution $\ha(\xx,t)$ from step (2), evaluate the time-dependent gradient $\mathrm{d} \OBJ_{t} / \mathrm{d} \cc$ with (\ref{eq:dJdc});
\item
Given the gradient $\mathrm{d} \OBJ_{t} / \mathrm{d} \cc$  from step (3), update the control $\cc_k(t) \to \cc_{k+1}(t)$ with a gradient-based method. Go back to step (1) and iterate until convergence.
\end{enumerate}

A few remarks are in order.
At the first iteration $k=1$, we initialize the control $(\theta(t),\beta(t))$  either with an arbitrary guess, or with one of the promising harmonic control laws obtained with the Monte-Carlo method. 
The choice of the initial control can have a substantial effect on the outcome of the optimization because $\OBJ_{t}(\cc)$ is not convex and may exhibit local minima, where gradients methods can stay trapped. 
To increase the chances of achieving a satisfactory reduction of $\OBJ_{t}$, we repeat the optimization from different initial guesses.

At each iteration, the full solution $h(\xx,t)$ is saved in step (1) and later used to solve the adjoint equation in step (2). 
Given the reasonable size of the solution (number of mesh points and number of time steps), we do not need to use intermediate checkpoints.

We update the control in step (4) with the Polak-Ribiere variant of the conjugate gradient method (for the choice of the descent direction) and Brent's method (for the choice of the descent distance in that direction).

Finally, the above steps are repeated until the relative variations of the two terms in $\OBJ_{t}$ are both smaller than $10^{-6}$.  
We observe that when this convergence criterion is satisfied,  the norm of the gradient $\mathrm{d} \OBJ_{t} / \mathrm{d} \cc$  is generally small too.

\subsection{Regulation control}
\label{sec:regulation}

The objective function (\ref{eq:J1_J2}) measures uniformity at the final time $t_f$ only (``terminal control''). 
It is  possible to use an alternative measure of uniformity that is distributed over the whole time interval $[0,t_f]$ (``regulation control''):
\begin{align}
\OBJ_{r} (h,\cc) &= \int_I \iint_D (h(\xx,t)-h_\text{opt})^2 \, \mathrm{d}\xx \, \mathrm{d}t
+ \gamma \intt \left( \theta(t)^2+\beta(t)^2 \right) \,\mathrm{d}t
\nonumber
\\
&= \int_I \mathcal{U}(t) \,  \mathrm{d}t
+ \gamma
\mathcal{C}.
\label{eq:J1_J2_regul}
\end{align}
Which formulation eventually leads to the best uniformity at $t=t_f$ is not obvious \textit{a priori}. 
It may be expected that terminal control, which  allows poorer short-term performance provided it yields a long-term advantage, is less restrictive and therefore more effective \cite{Bewley2001}.
In this study we focus on terminal control, but for the sake of completeness we nonetheless consider the optimization problem 
\begin{align}
\min_{\cc(t)} \OBJ_{r}(h,\cc)
\quad 
\mbox{ subject to (\ref{eq:governing}),}
\label{eq:constrained_opt_pb_reg}
\end{align}
with the same initial and boundary conditions as for (\ref{eq:constrained_opt_pb}).
The  Lagrangian is 
\begin{align}
\LAG_r(h,\cc,\ha)
=&
\OBJ_{r}(h,\cc) 
- 
\intt \iint_D \ha \left(\dfrac{\partial h}{\partial t} + \bnabla \bcdot \qq \right) \,\mathrm{d}\xx \,\mathrm{d}t
\\
=&
\int_I \iint_D (h(\xx,t_f)-h_\text{opt})^2 \, \mathrm{d}\xx\,\mathrm{d}t
+ \gamma \intt \left( \theta(t)^2+\beta(t)^2 \right) \,\mathrm{d}t
\nonumber
\\
&
-\intt \iint_D \ha \left(\dfrac{\partial h}{\partial t} + \bnabla \bcdot \qq \right) \,\mathrm{d}\xx \,\mathrm{d}t,
\end{align}
and the adjoint equation associated with this regulation control problem reads:
\begin{align}
 \dfrac{\partial \ha}{\partial t}
+ W  
\bnabla  \bcdot \left( h^3  \cost \cosb \bnabla \ha  \right) 
-3 W h^2 
 \left( \cost \cosb \bnabla h - \SSS \right) 
 \bcdot \bnabla\ha
\nonumber
\\ 
= -2(h-h_\text{opt}),
\label{eq:adj_regul}
\end{align}
with terminal and boundary conditions
\begin{align}
& \ha(\xx,t_f) = 0,
\label{eq:adj_init_regul}
\\
& \bnabla \ha \bcdot \nn_D = 0 
\quad \mbox{ on }\partial D.
\label{eq:adj_BC_regul}
\end{align}
In contrast to the terminal control problem, where the adjoint equation (\ref{eq:adj}) is homogeneous and the adjoint dynamics are determined by the non-zero terminal condition (\ref{eq:adj_init}),
in the regulation control problem the terminal condition (\ref{eq:adj_init_regul}) is zero and the adjoint dynamics are determined by the time-dependent forcing on the right-hand side of (\ref{eq:adj_regul}). 
Finally, the gradient needed to update the control so as to decrease $\OBJ_{r}$ reads
\begin{align}
\dfrac{\mathrm{d} \OBJ_{r}}{\mathrm{d}\cc}
&= 
2\gamma 
\left(
\begin{array}{c} \theta(t) \\ \beta(t) 
\end{array} \right)
- \iint_D \ha 
\left(
\dfrac{\partial \left(\bnabla \bcdot \qq\right)}{\partial \theta},
\dfrac{\partial \left(\bnabla \bcdot \qq\right)}{\partial \beta} 
\right)^T
\,\mathrm{d}\xx.
\label{eq:dJdc_regul}
\end{align}

\section{Optimal control results}
\label{sec:opt_control_res}

\begin{figure}[]
\hspace{-0.3cm} 
\subfigure[]{
  \includegraphics[trim={15mm 62mm 18mm 65mm}, clip, height=6.5cm]{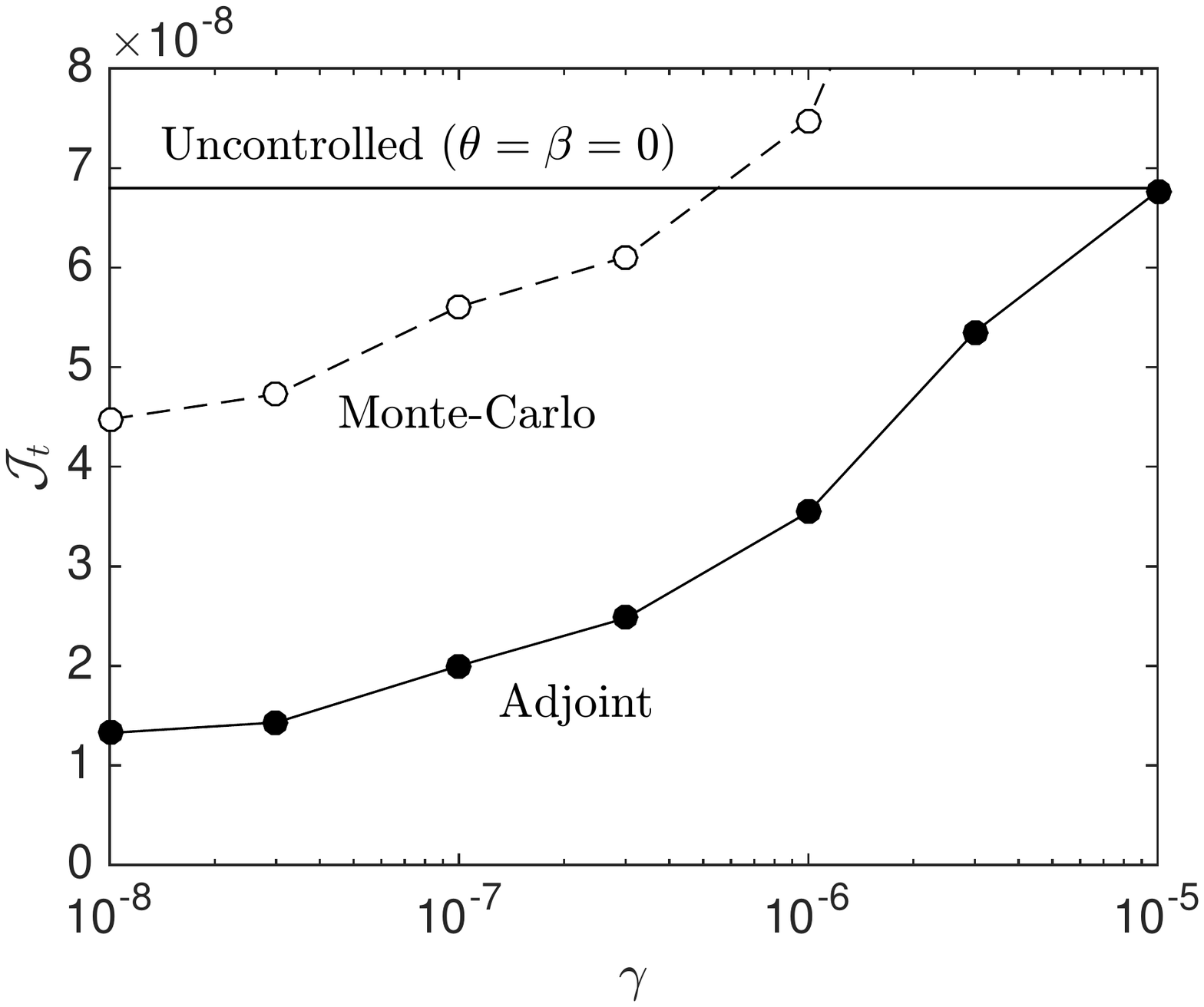}
} 
\subfigure[]{
   \includegraphics[trim={15mm 62mm 18mm 65mm}, clip, height=6.5cm]{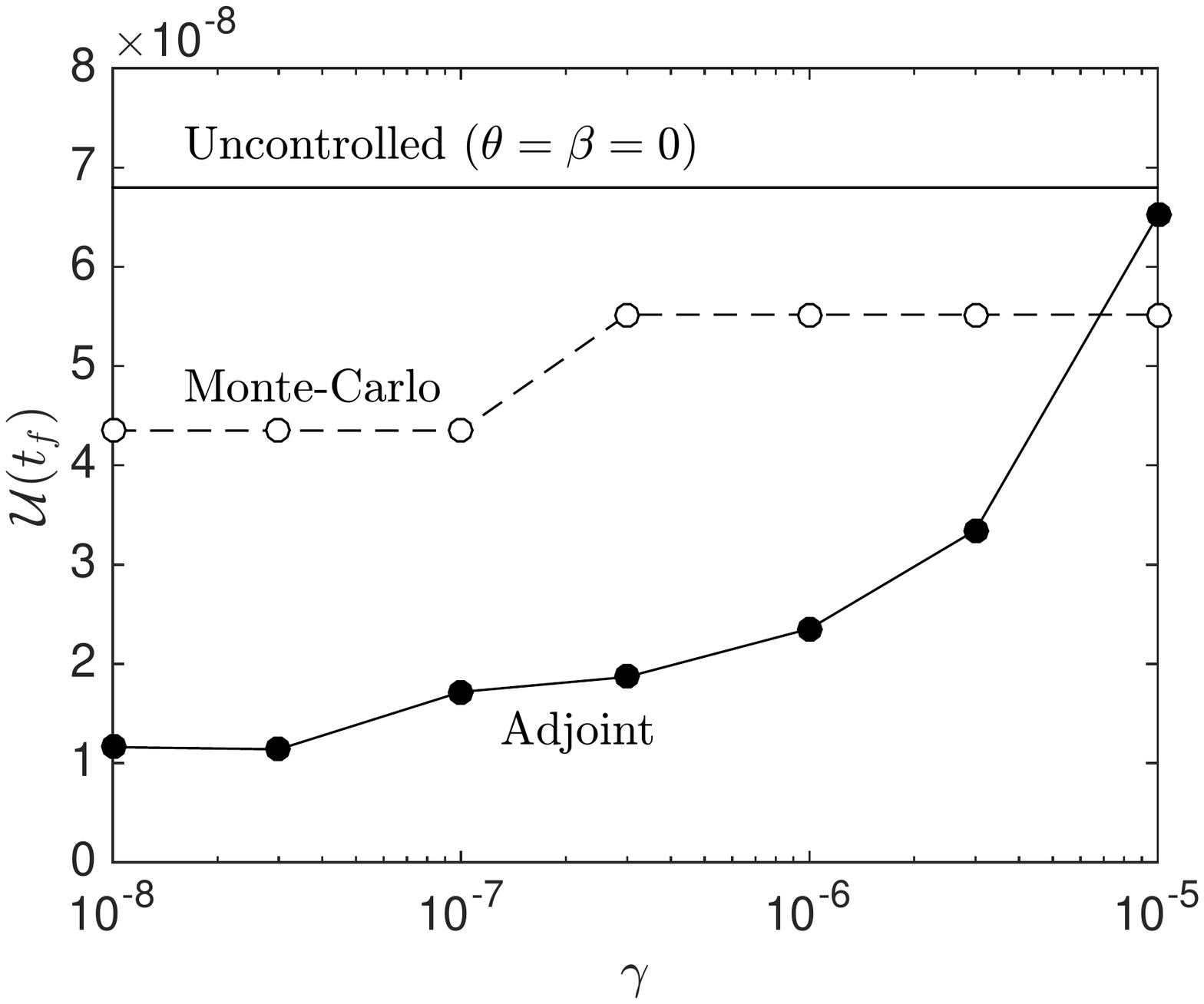}
}
\caption{(a)~Objective function $\OBJ_{t}$ and (b)~corresponding final uniformity  $\mathcal{U}(t_f)$, vs. unit cost of the control $\gamma$.}
\label{fig:J_J1}
\end{figure}

Results of the optimization are reported in Fig.~\ref{fig:J_J1} for a wide range of unit cost values $\gamma \in [10^{-8}, 10^{-5}]$. 
Fig.~\ref{fig:J_J1}(a) shows the  objective function $\OBJ_{t}$, and Fig.~\ref{fig:J_J1}(b) the final uniformity measure $\mathcal{U}(t_f)$.
(In this section, we give numerical values of $\mathcal{U}(t_f)$ in m$^4$ and omit units.)
Recall that, as defined in (\ref{eq:J1_J2}), small $\gamma$ values correspond to cheap controls (large angles $|\theta|$, $|\beta|$  are not penalized as they barely affect $\OBJ_{t}$), whereas large $\gamma$ values correspond to expensive controls (large angles $|\theta|$, $|\beta|$ are penalized as they contribute to increasing $\OBJ_{t}$).

The uncontrolled case (fixed horizontal substrate, $\theta=\beta=0$) leads at $t=t_f$ to a uniformity $\mathcal{U}(t_f) = 6.80\times 10^{-8}$  (solid line; see also section~\ref{sec:fixed_substrate}).
Regarding time-harmonic control obtained with the Monte-Carlo algorithm  (dashed line with open symbols; see also section~\ref{sec:Monte_Carlo}), the best control for $\OBJ_t$ yields $\mathcal{U}(t_f) = 4.35\times 10^{-8}$ for small $\gamma$ values. 
As $\gamma$ increases,  the second-best control in terms of uniformity ($\mathcal{U}(t_f) = 5.52\times 10^{-8}$) becomes the best control for $\OBJ_t$  thanks to its smaller angles and lower total cost.
However, although this time-harmonic control yields a slightly better uniformity $\mathcal{U}(t_f)$ than the uncontrolled case, it eventually becomes less efficient for $\OBJ_t$ when $\gamma \gtrsim 5\times 10^{-7}$.

The adjoint optimization (filled symbols) yields a substantial improvement. 
For small $\gamma$, the uniformity is improved by a factor 6,
down to $\mathcal{U}(t_f) = 1.16\times 10^{-8}$. 
As the unit cost increases, $\OBJ_t$ and $\mathcal{U}(t_f)$ increase but remain smaller than their uncontrolled counterparts up to $\gamma$ as large as $10^{-5}$.

\begin{figure}[]
\begin{center}
\subfigure[]{
  \includegraphics[trim={10mm 65mm 22mm 70mm}, clip, height=5.8cm]{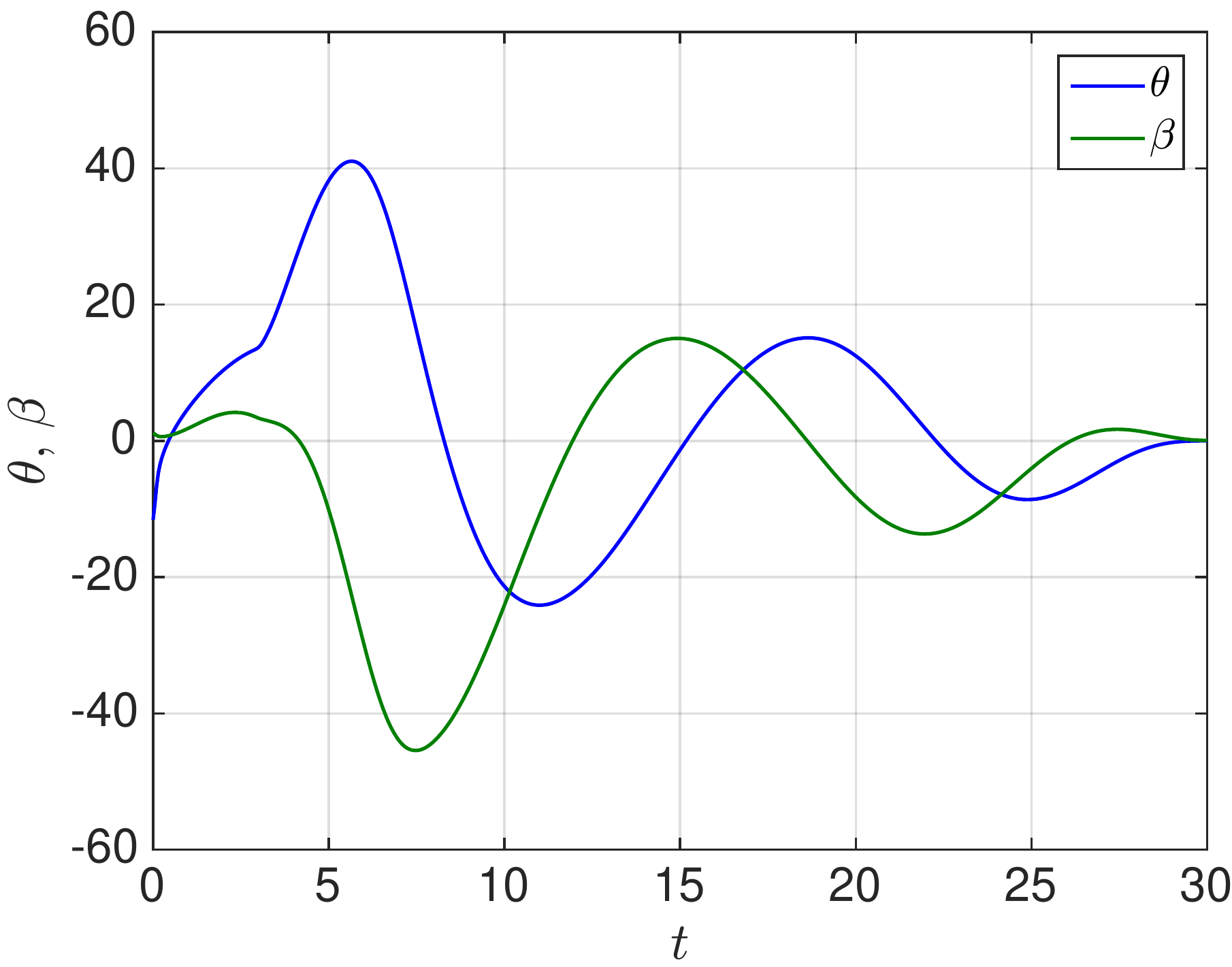}
}
\subfigure[]{
  \includegraphics[trim={10mm 65mm 22mm 70mm}, clip, height=5.8cm]{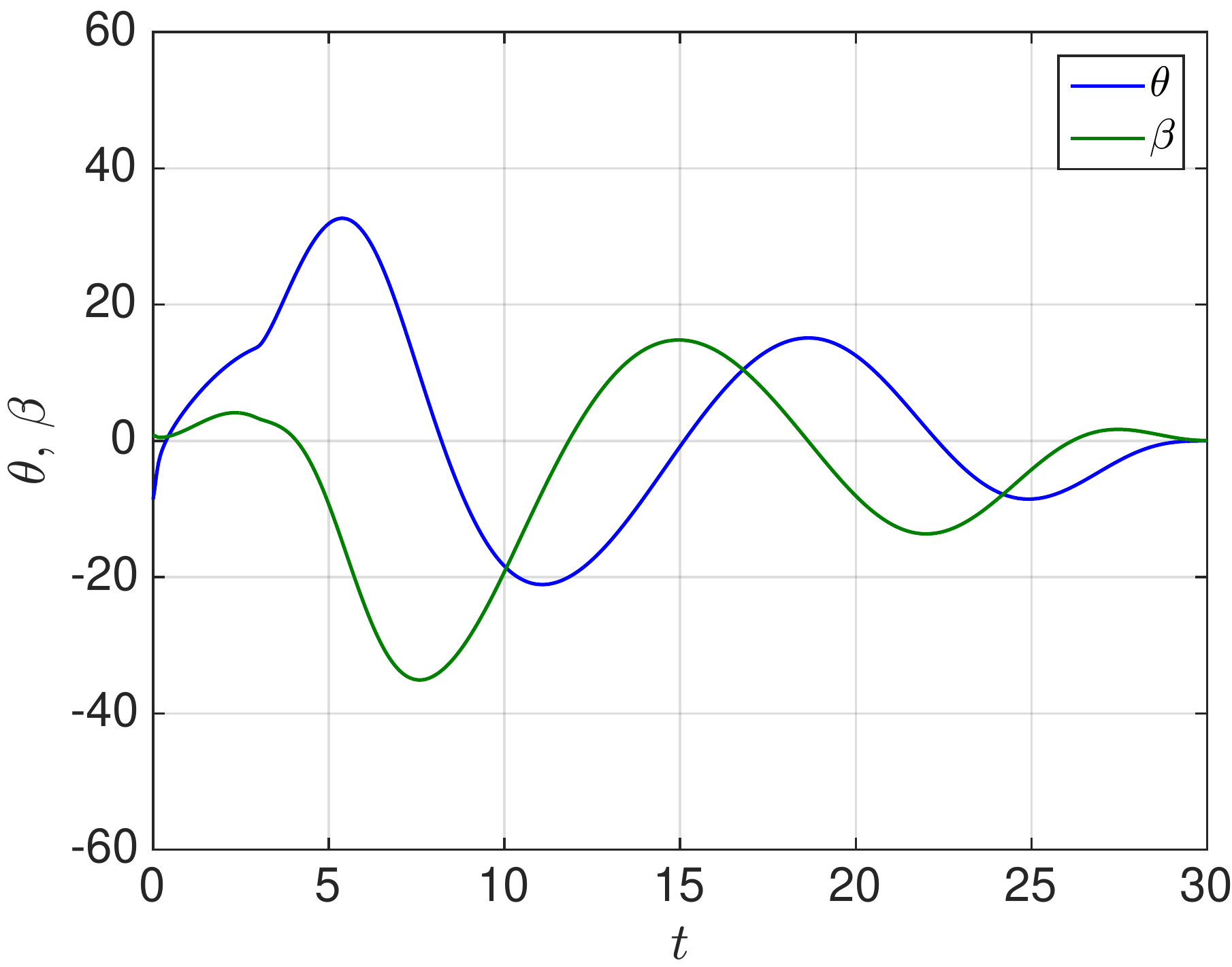}
}
\end{center}
\begin{center}
\subfigure[]{
  \includegraphics[trim={10mm 65mm 22mm 70mm}, clip, height=5.8cm]{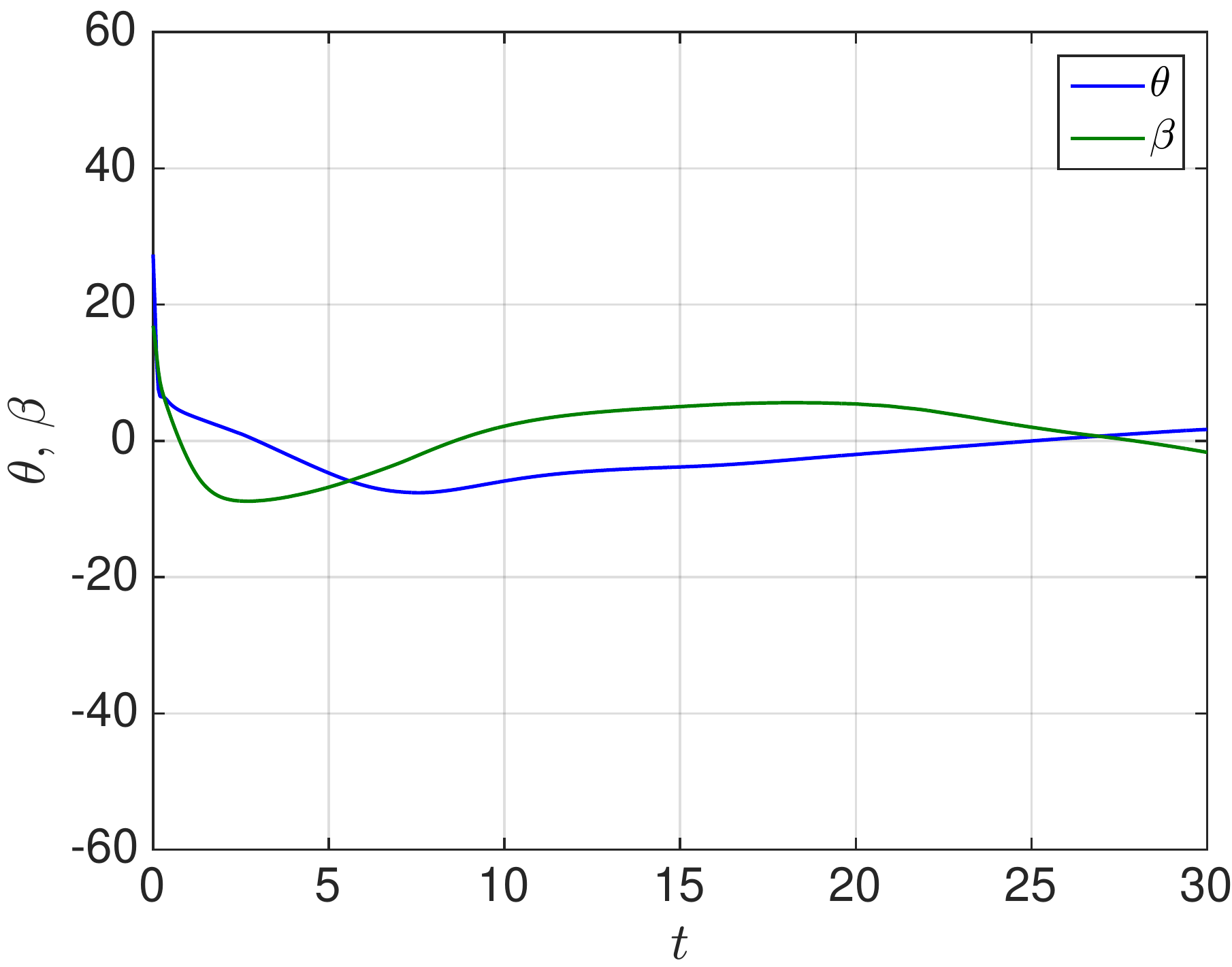}
}
\subfigure[]{
  \includegraphics[trim={10mm 65mm 22mm 70mm}, clip, height=5.8cm]{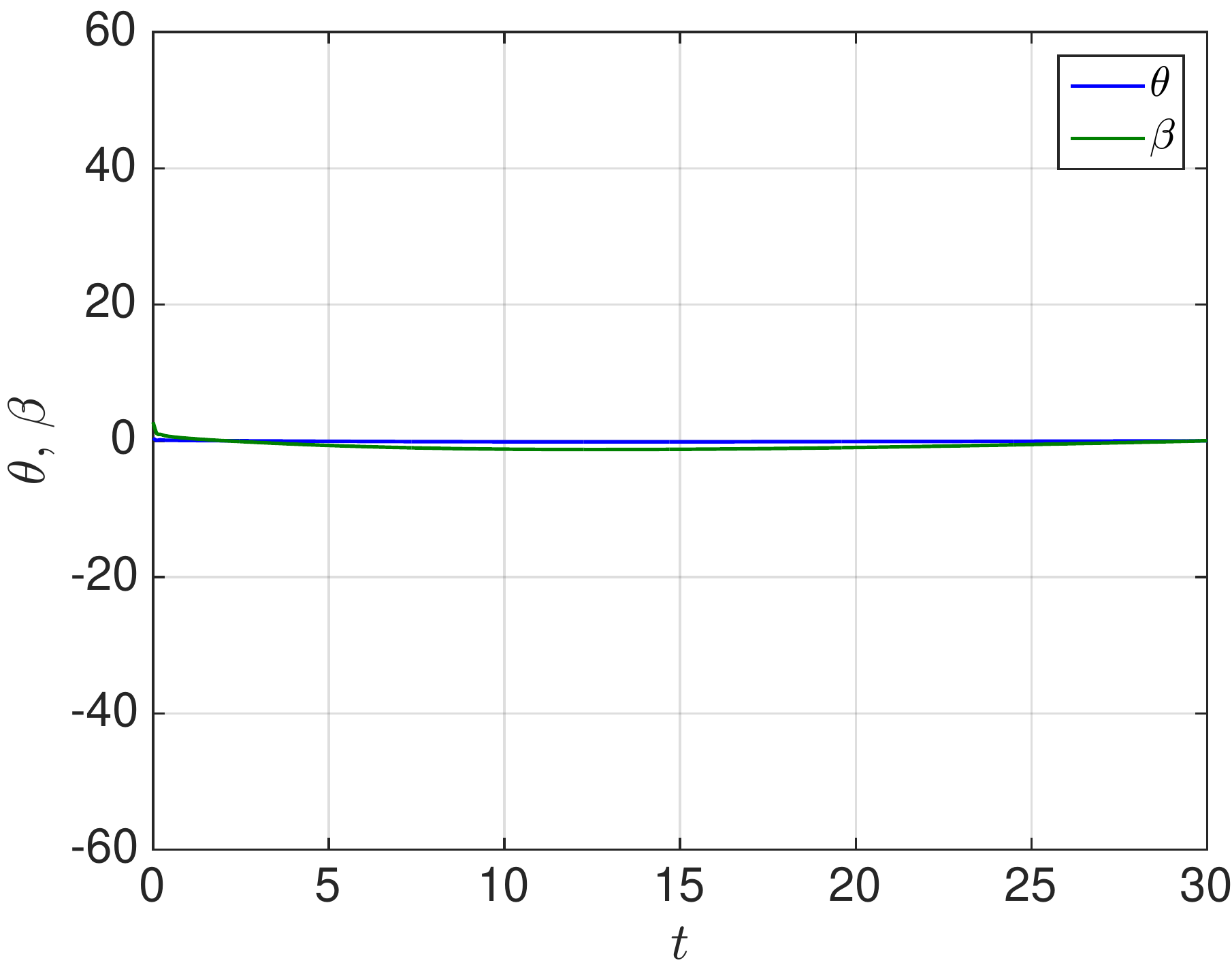}
}
\end{center}
\caption{ 
Optimal control $\theta(t)$, $\beta(t)$ obtained with the adjoint optimization, for different values  of the unit cost of the control:
(a)~$\gamma=10^{-8}$,
(b)~$\gamma=10^{-7}$,
(c)~$\gamma=10^{-6}$,
(d)~$\gamma=10^{-5}$.
}
\label{fig:curves_opt_ctrl_adjoint}
\end{figure}

Figure~\ref{fig:curves_opt_ctrl_adjoint} shows the best control laws  $\theta(t)$, $\beta(t)$ obtained with the adjoint optimization. 
Clearly, a larger unit cost  leads to smaller angles: while inclinations as large as $45^o$ are admissible for $\gamma=10^{-8}$, they do not exceed $10^o$ for $\gamma=10^{-6},$ and barely differ from the no-control case $\theta=\beta=0$ for $\gamma=10^{-5}$.
The kinematics displayed here is close to a damped rotation motion: $\theta(t)$ and $\beta(t)$ are approximately periodic, decreasing in amplitude, and in phase quadrature (azimuthally traveling wave).
About two rotations are described for $\gamma \leq 10^{-7}$ and  one rotation for $\gamma \leq 10^{-6}$, possibly because the larger angles allowed for small $\gamma$ can spread the film faster, which in turn allows for faster kinematics.

This is confirmed by the time evolution of the film thickness in Figs.~\ref{fig:Adj_1e-8}-\ref{fig:Adj_1e-6} (see also videos in the supplementary material \cite{SM}). 
For $\gamma = 10^{-8}$ (Fig.~\ref{fig:Adj_1e-8}), first the bulk of the liquid is  quickly moved to the rim of the disk, leaving behind a thinner film in the central region of the disk ($0 \leq t \leq \Delta t$); 
the thickest portion of the film is then displaced along the whole rim, thus depositing some liquid in the yet uncovered disk areas (first rotation, $\Delta t \leq t \leq 4 \Delta t$); 
finally, another rotation further distributes the liquid and improves the thickness uniformity, albeit less markedly because the film has already 
become more viscous.
For $\gamma = 10^{-6}$ (Fig.~\ref{fig:Adj_1e-6}), smaller inclinations cause the fluid to move more slowly.
Only about one rotation is completed by the time
 viscosity becomes so large
that the thickness uniformity cannot be modified substantially. 
However, the optimized control manages to distribute the liquid just about everywhere over the disk. 
Remarkably, this is achieved by gradually slowing down the rotating motion as the film 
becomes more viscous (compare 
$0          \leq t \leq 2 \Delta t$ and  
$2 \Delta t \leq t \leq 6 \Delta t$ in Figs.~\ref{fig:curves_opt_ctrl_adjoint}(c) and \ref{fig:Adj_1e-6}).

\begin{figure}
\centering
\includegraphics[scale=1]{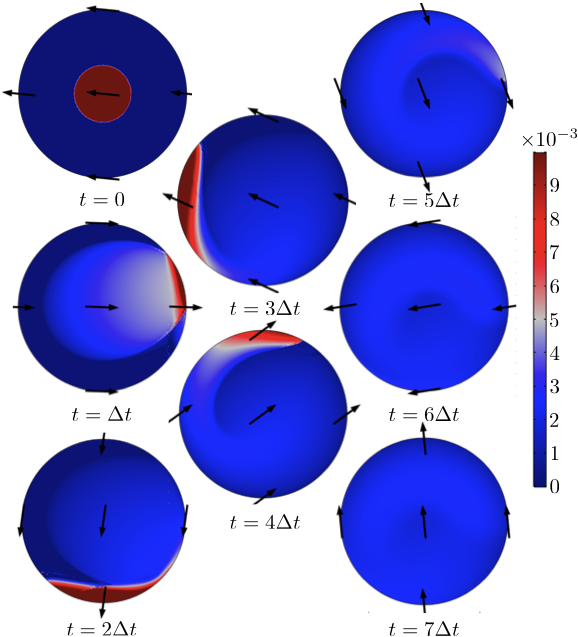}
\caption{
Contours of film thickness $h(\xx,t)$ for the best adjoint control for $\gamma = 10^{-8}$.
Same representation as Figs.~\ref{fig:MC1}-\ref{fig:MC2}.
See video 3 in the supplementary material \cite{SM}.
}
\label{fig:Adj_1e-8}
\end{figure}

\begin{figure}
\centering
\includegraphics[scale=1]{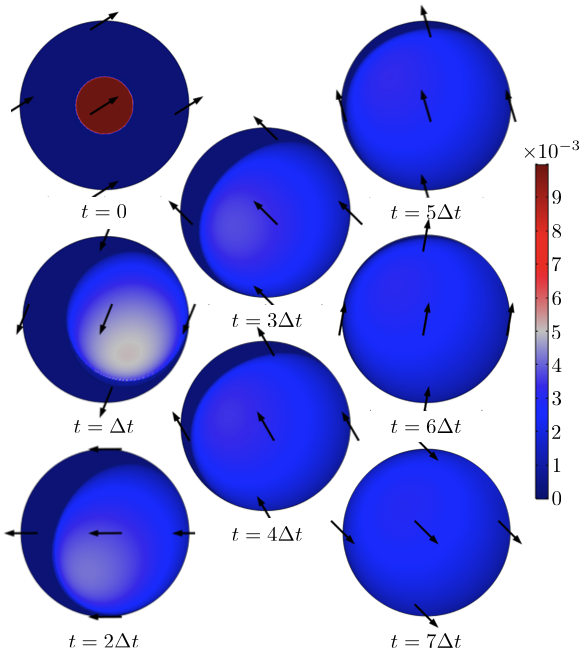}
\caption{
Contours of film thickness $h(\xx,t)$ for the best adjoint control for $\gamma = 10^{-6}$.
Same representation as Figs.~\ref{fig:MC1}-\ref{fig:MC2}.
See video 4 in the supplementary material \cite{SM}.
}
\label{fig:Adj_1e-6}
\end{figure}

\begin{figure}[]
\begin{center}
\includegraphics[scale=1]{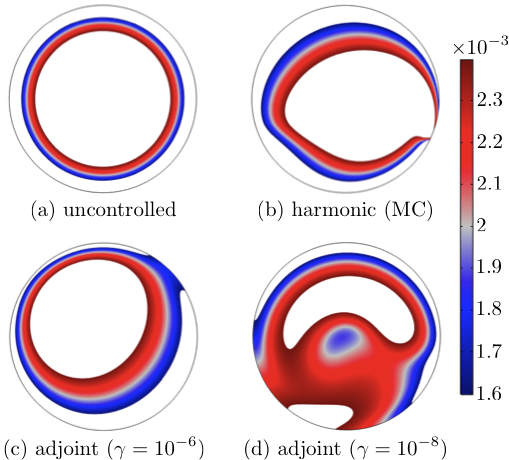}
\end{center}
\vspace{-0.6cm}
\caption{ 
Final thickness $h(\xx,t_f)$ clipped between $0.8h_\text{opt}$ and $1.2h_\text{opt}$ for 
(a)~the uncontrolled case; 
(b)~the optimal harmonic control;
(c)~the optimal adjoint control with $\gamma = 10^{-6}$; 
(d)~the optimal adjoint control with $\gamma = 10^{-8}$.
From (a) to (d), the final film thickness is close to the optimal uniform thickness $h_\text{opt}$ over an increasingly wide area.
}
\label{fig:Final_contours}
\end{figure}

For a better visualization of the final state,
Fig.~\ref{fig:Final_contours} shows contours of film thickness in regions where it is within 20\% of $h_\text{opt}$.
Without control, $h(\xx,t_f)$ falls in this interval only in a narrow ring: the inner region of the substrate remains much thicker than $h_\text{opt}$, and the outer region remains much thinner, resulting in a highly non-uniform film.
Control improves the film spreading, albeit in a non-axisymmetric fashion.
The best harmonic control, expensive adjoint control, and cheap adjoint control result in an increasingly wider area where the film thickness departs from $h_\text{opt}$ by less than 20\%
(see also Figs.~\ref{fig:MC1} and \ref{fig:Adj_1e-8}-\ref{fig:Adj_1e-6}).

\begin{figure}[htp]
\centering
\includegraphics[trim={8mm 65mm 22mm 70mm}, clip, height=5.8cm]{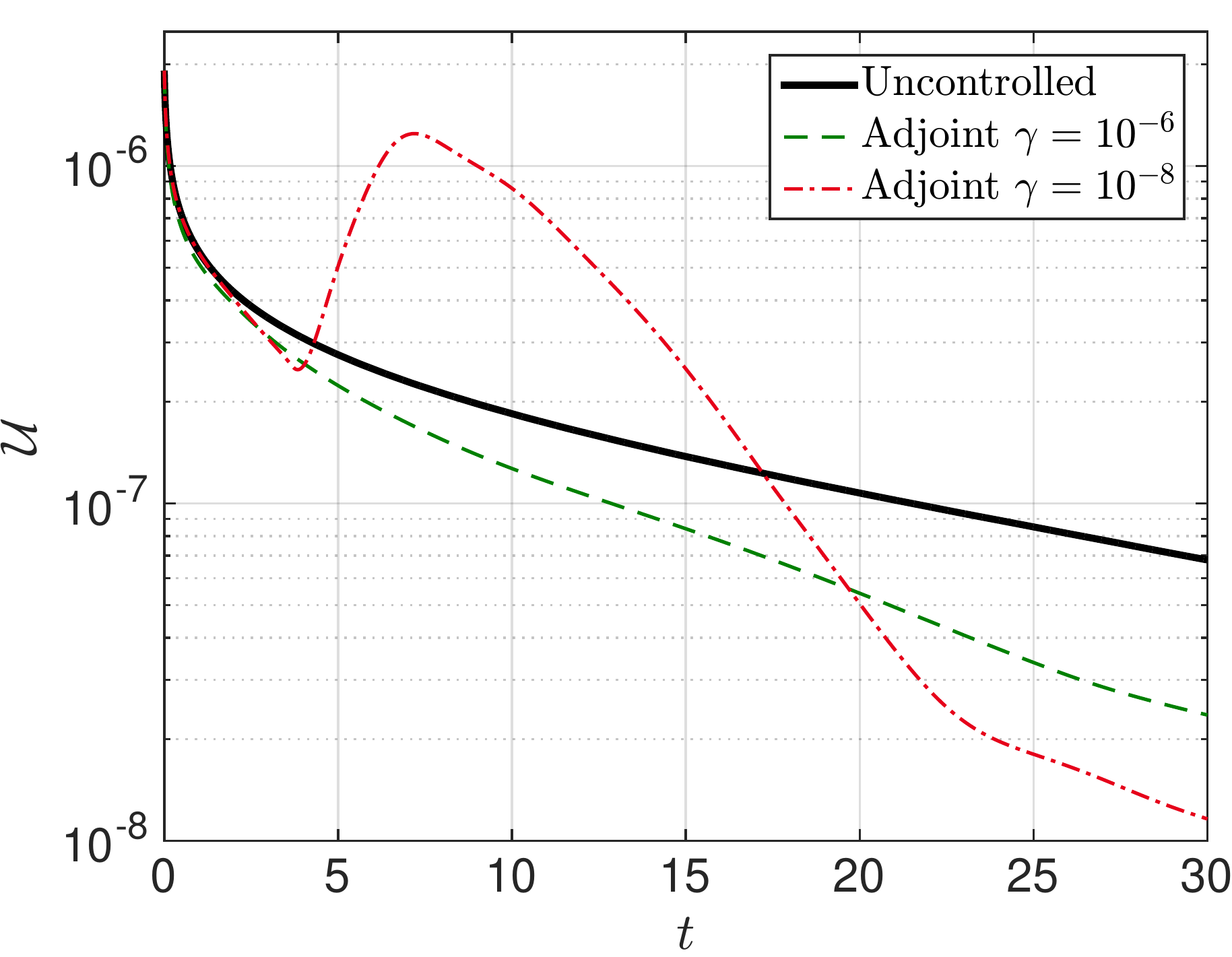}
\caption{
Time evolution of the measure $\mathcal{U}(t)$ of film thickness uniformity.
No control vs. optimal adjoint control for $\gamma=10^{-6}$ and $\gamma=10^{-8}$. 
}
\label{fig:U_vs_t-adjoint}
\end{figure}

The evolution of $\mathcal{U}(t)$ in Fig.~\ref{fig:U_vs_t-adjoint}
shows interesting features.
First, as already mentioned, the adjoint optimization improves the final thickness uniformity   compared to the uncontrolled case, and more substantially as  $\gamma$ decreases.
Second, our terminal control formulation allows  $\mathcal{U}(t)$ to \textit{increase temporarily}  because this yields a \textit{lower final}  value $\mathcal{U}(t_f)$.
In the specific case at hand ($\gamma=10^{-8}$), a sharp increase around $t\simeq5$~s brings $\mathcal{U}$ back to its initial (and maximal) value. 
With less time available and a liquid that has become more viscous, the situation may seem compromised. 
However, from this point onward,  $\mathcal{U}$ decreases quickly and steadily, finally yielding the best uniformity achieved in the present study.

\begin{figure}[]
\begin{center}
\includegraphics[scale=1]{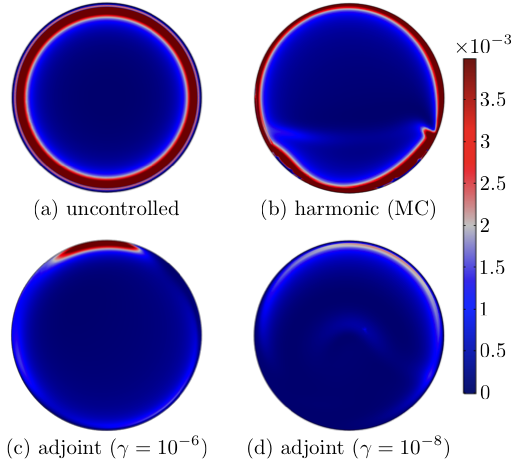}
\end{center}
\vspace{-0.6cm}
\caption{ 
Final thickness gradient squared $\left| \bnabla h(\xx,t_f) \right|^2$ for 
(a)~the uncontrolled case; 
(b)~the optimal harmonic control; 
(c)~the optimal adjoint control with $\gamma = 10^{-6}$; 
(d)~the optimal adjoint control with $\gamma = 10^{-8}$.
From (a) to (d),  the final film  is increasingly smoother (i.e. with smaller gradients).
}
\label{fig:Final_grad}
\end{figure}

One may wonder whether more uniform films are also smoother. While of course this is not true in general, it is worth investigating the smoothness of the final films obtained with and without control.
A simple measure of smoothness is based on the gradient of the thickness:
\be 
\mathcal{S}(t) = \iint_{D} \left| \bnabla h(\xx,t) \right|^2 \text{d}\xx.
\label{eq:smoothness}
\ee 
A uniform film yields $\mathcal{S}=0$, while large gradients $\bnabla h$ yield large values of $\mathcal{S}$.
Other measures exist, putting for instance  different weights on small or large spatial wavelengths \cite{Bewley2001, Foures2014}, but (\ref{eq:smoothness}) suffices for present purposes.
In all cases, we observe that $\mathcal{S}(t_f)<\mathcal{S}(0)$: overall, the film becomes smoother as time evolves. 
This is consistent with the initial film exhibiting a localized but very sharp front at $r=R_i$, whereas final films have smoother fronts due to the effect of gravity.
We also note that film uniformity and smoothness are indeed correlated.
Table~\ref{Tab:Utf_Stf} summarizes final values of $\mathcal{U}$ and $\mathcal{S}$.
Both measures yield similar trends in terms of control performance:  cheaper adjoint control, followed by more expensive adjoint control,  best harmonic controls, and no control.
This trend is also illustrated by the final gradient fields in Fig.~\ref{fig:Final_grad}.
This suggests that good results in terms of uniformity $\mathcal{U}$ could be obtained by formulating the optimization with the smoothness measure $\mathcal{S}$ in the objective function.
In some cases, it so happens that optimization problems formulated with an objective function different from the true objective yield better result (see~\cite{Bewley2001} for an example where drag reduction in a plane channel flow is best achieved by targeting kinetic turbulent energy).

\begin{table}
\begin{center}
  \begin{tabular}{|l|c|c|}
    \hline
    & Final uniformity $\mathcal{U}(t_f)$ & Final smoothness $\mathcal{S}(t_f)$ 
    \\
    & $\iint_{D}\left( h(\xx,t_f)-h_\text{opt} \right)^2 \text{d}\xx$ 
    & $\iint_{D} \left| \bnabla h(\xx,t_f) \right|^2 \text{d}\xx$ 
    \\
    & (m$^4$) 
    & (m$^2$) 
    \\ \hline
    Uncontrolled & 
    $6.8 \times 10^{-8}$ & 
    $3.6 \times 10^{-4}$ 
    \\ 
    Best harmonic (MC) & 
    $5.5 \times 10^{-8}$ & 
    $1.4 \times 10^{-4}$ 
    \\ 
    2$^{nd}$ best harmonic (MC) & 
    $4.3 \times 10^{-8}$ & 
    $1.6 \times 10^{-4}$  
    \\ 
    Adjoint $\gamma=10^{-6}$ & 
    $2.4 \times 10^{-8}$ & 
    $0.31 \times 10^{-4}$ 
    \\ 
    Adjoint $\gamma=10^{-8}$ & 
    $1.2 \times 10^{-8}$ & 
    $0.20 \times 10^{-4}$ 
    \\ \hline
  \end{tabular}
\caption{
Final uniformity and smoothness. Improvements in uniformity and smoothness follow the same trend.
}
\label{Tab:Utf_Stf}
\end{center}
\end{table}

Interestingly, we have obtained other controls, with qualitatively different kinematics but similar values of $\OBJ_t$ and $\mathcal{U}(t_f)$. 
This suggests that the objective function has several local minima, as mentioned in section~\ref{sec:adjoint_sol_proc}.
We cannot rule out the possibility that better controls exist. 
However, running the optimization algorithm from a set of initial guesses did allow us to decrease $\mathcal{U}(t_f)$ substantially compared to both the uncontrolled case and the best time-harmonic Monte-Carlo controls, with a set of physical parameters  constituting a challenging test case, as cooling occurs quickly and the fluid becomes more viscous.

The adjoint optimization algorithm generally converged within 10 to 30 gradient evaluations, each iteration consisting of 1 adjoint calculation (to determine the descent direction) and 5 to 10 direct calculations (to determine the descent distance in that direction). This is to be contrasted with the 2000 evaluations and  poorer results of the Monte-Carlo algorithm. 
(To be fair, it must be recalled that the Monte-Carlo algorithm was restricted to harmonic controls, that are necessarily sub-optimal compared to arbitrary controls.)
It is precisely the very large number of degrees of freedom, optimized efficiently within a reasonably small number of iterations, that makes adjoint methods powerful and attractive for optimal control in this and other settings.

\section{Conclusions}
\label{sec:disc_conc}

The question at the center of this study is: can one improve the coverage of a surface by a gravity-driven liquid film using a suitable  kinematics of the surface to distribute the liquid uniformly? 
This question is answered in the framework of PDE-constrained optimization, for which ones seeks to minimize an objective function (a combination of the film thickness uniformity and of the cost of moving the substrate) subject to a set of constraints (governing PDE with associated boundary and initial conditions). 
The PDE that governs the film dynamics is obtained in the lubrication approximation framework, using a temperature-dependent viscosity and a gravity force whose direction and magnitude depend on the orientation of the surface. 
This effectively expresses the surface kinematics as a time-dependent body force in the governing equations. 
This model has inherent limitations, namely that the flow must be free of inertia, and the free surface slope must be small. 
Both assumptions appear to be reasonable for the applications of interest here: pancake making and surface coating. 

Two methods are proposed to optimize the film uniformity. 
The first one assumes a kinematics described by harmonic functions, and identifies the optimal amplitudes and periods from a set of 
randomly distributed Monte-Carlo realizations. 
This method is computationally costly as it blindly samples the large parameter space, and it is shown to offer some improvement in thickness uniformity (40\% improvement compared to the uncontrolled case). 
The second method is a gradient-based method that allows for arbitrary controls.
We have shown  that the gradient of the objective function can be conveniently calculated using an adjoint formulation. 
This second approach is shown to be much more effective as it leads to a 83\% improvement of the uniformity measure for  the cheapest control ($\gamma = 10^{-8}$). 
The computational cost of this second method is lower than that of the Monte-Carlo method since local information is available to guide the optimization in a suitable  descent direction. 
Interestingly, one of the optimal controls identified by the adjoint method appears to replicate the motion one would naturally adopt when making cr\^epes, i.e. first draining all the batter to one end of the pan, and then slowly rotating it for one revolution or two in order to distribute the batter in the remaining part of the pan. 
Results also suggest the existence of multiple local minima since different kinematics can lead to commensurate uniformity. 
The framework presented here can be adapted to other optimal control problems involving thin liquid films.

\section{Acknowledgments}

The authors wish to express their gratitude to D. Sellier for carefully proofreading the manuscript.
The authors also thank the financial support of the University of Canterbury through the College of Engineering Strategic Research grant and the Innovation Jumpstart grant.

\bibliographystyle{apsrev}

\end{document}